\documentclass{aa} 
\usepackage{amsmath} 
\usepackage{txfonts}
\usepackage{graphicx} 
\usepackage{natbib} 
\bibpunct{(}{)}{;}{a}{}{,}

\newcommand{\diff}{\mathrm{d}}

\newcommand{\ie}{\textit{i.e.},}
\newcommand{\eg}{\textit{e.g.},}
\newcommand{\Ms}{\ensuremath{M_\odot}}

\newcommand{\el}[2]{$\rm{}^{#2}\kern-0.6pt#1$}

\newcommand{\fll}[1]{\ensuremath{f_\text{LL}^\text{#1G}}}

\newcommand{\frem}[1]{\ensuremath{f_\text{rem}^\text{#1G}}}

\newcommand{\elm}[1]{\ensuremath{e_\text{LL}^\text{#1G}}}
\newcommand{\erem}[1]{\ensuremath{e_\text{rem}^\text{#1G}}}

\begin{document}

\title{Origin of the abundance patterns in Galactic globular clusters: constraints on dynamical and chemical properties of globular clusters}


\author{T. Decressin\inst{1} \and C. Charbonnel\inst{1,2}\and G. Meynet\inst{1} }

\offprints{T. Decressin, \\Thibaut.Decressin@obs.unige.ch}

\institute{Observatoire de Gen\`eve, Universit\'e de Gen\`eve, 51, ch. des
  Maillettes, 1290 Sauverny, Switzerland \and Laboratoire d'Astrophysique
  de Toulouse et Tarbes, CNRS UMR 5572, OMP, Universit\'e Paul Sabatier Toulouse
  3, 14, Av. E. Belin, 31400 Toulouse, France }

\date{Received / Accepted}

\authorrunning{} \titlerunning{}

\abstract{} %
{We analyse the effects of a first generation of fast rotating massive
  stars on the dynamical and chemical properties of globular clusters.}%
{We use stellar models of fast rotating massive stars, losing mass through
  a slow mechanical equatorial winds to produce material rich in H-burning
  products. We propose that stars with high Na and low O abundances
  (hereafter anomalous stars) are formed from matter made of slow winds of
  individual massive stars and of interstellar matter. The proportion of
  slow wind and of interstellar material is fixed in order to reproduce the
  observed Li-Na anticorrelation in NGC 6752.}%
{In the case that globular clusters, during their lifetime, did not lose
  any stars, we found that to reproduce the observed ratio of normal to
  anomalous stars, a flat initial mass function (IMF) is needed, with
  typically a slope $x=0.55$ (a Salpeter's IMF has $x=1.35$). In the case that
  globular clusters suffer from an evaporation of normal stars, the IMF
  slope can be steeper: to have $x=1.35$, about 96\% of the normal stars
  would be lost. We make predictions for the distribution of stars as a
  function of their [O/Na] and obtain quite reasonable agreement with that
  one observed for NGC 6752. Predictions for the number fraction of stars
  with different values of helium, of the \el{C}{12}/\el{C}{13} and
  \el{O}{16}/\el{O}{17} ratios are discussed, as well as the expected
  relations between values of [O/Na] and those of helium, of [C/N], of
  \el{C}{12}/\el{C}{13} and of \el{O}{16}/\el{O}{17}. Future observations
  might test these predictions. We also provide predictions for the present
  day mass of the clusters expressed in units of mass of the gas used to
  form stars, and for the way the present day mass is distributed between
  the first and second generation of stars and the stellar remnants.}%
{}


\maketitle

\section{Introduction}

It has long been known that globular cluster stars present some striking
anomalies in their content in light elements\footnote{On the contrary, the
  content in heavy elements (i.e., Fe-group, $\alpha$-elements) is fairly
  constant from star to star in any well-studied individual Galactic
  globular cluster (with the notable exception of $\omega$~Cen).}: while in
all the Galactic globular clusters studied so far one finds ``normal'' stars
with detailed chemical composition similar to those of field stars of same
metallicity (i.e., same [Fe/H]), one also observes numerous ``anomalous''
main sequence and red giant stars that are simultaneously deficient (to
various degrees) in C, O, and Mg, and enriched in N, Na, and Al (for recent
reviews see \citealp{GrattonSneden2004,Charbonnel2005}).
Additionally, the abundance of the fragile Li was found to be
anticorrelated with that of Na in turnoff stars in a couple of globular
clusters (\citealp{PasquiniBonifacio2005,BonifacioPasquini2007}).

It is clear now that these chemical peculiarities were inherited at
birth by the low-mass stars we observe today in globular clusters, and that
their root cause is H-burning through the CNO-cycle and the NeNa- and
MgAl-chains that occurred in an early generation of more massive and faster
evolving globular cluster stars\footnote{As in field stars, the surface
  abundances of Li, C, and N also vary in globular cluster stars due to in
  situ evolutionary processes (i.e., first dredge-up and thermohaline
  instability on the red giant branch; see Charbonnel \& Zahn 2007 and
  references therein). We will not address this point in the present
  paper.}  (see \citealp{PrantzosCharbonnel2006}, hereafter
\defcitealias{PrantzosCharbonnel2006}{PC06}\citetalias{PrantzosCharbonnel2006},
and references therein). In other words, compelling evidence leads us to
believe that at least two generations of stars succeeded in all Galactic
globular clusters during their infancy. The first one corresponds to the
bulk of ``normal'' stars born with the pristine composition of the
protocluster gas; these objects are those with the highest O and Mg and the
lowest Na and Al abundances also found in their field contemporaries.  The
second generation contains the stars born out of material polluted to
various degrees by the ejecta of more massive stars, and which present
lower O and Mg and higher Na and Al abundances than their first generation
counterparts.

\begin{figure}[tbp]
  \centering
  \includegraphics[width=0.5\textwidth]{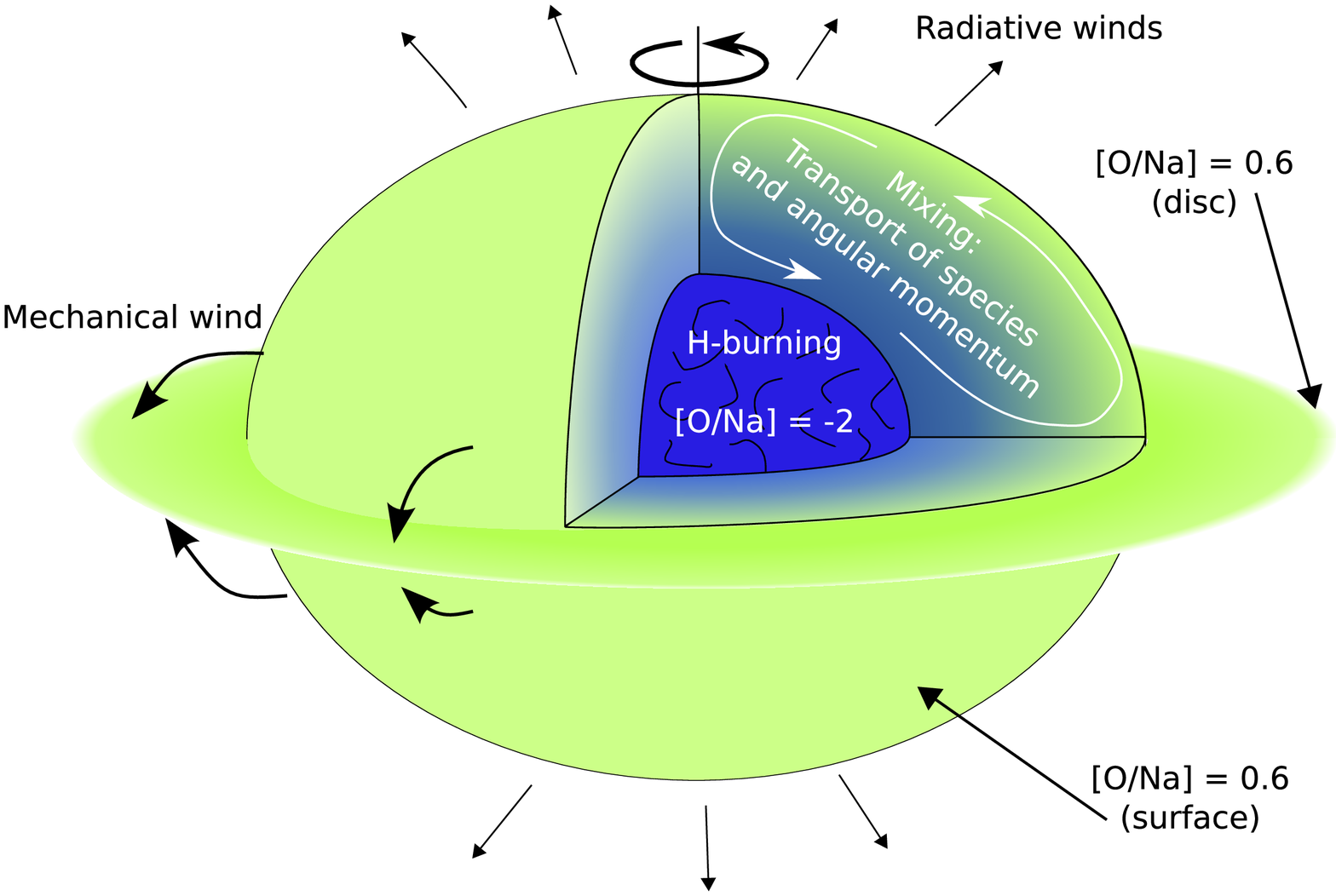}
  \includegraphics[width=0.5\textwidth]{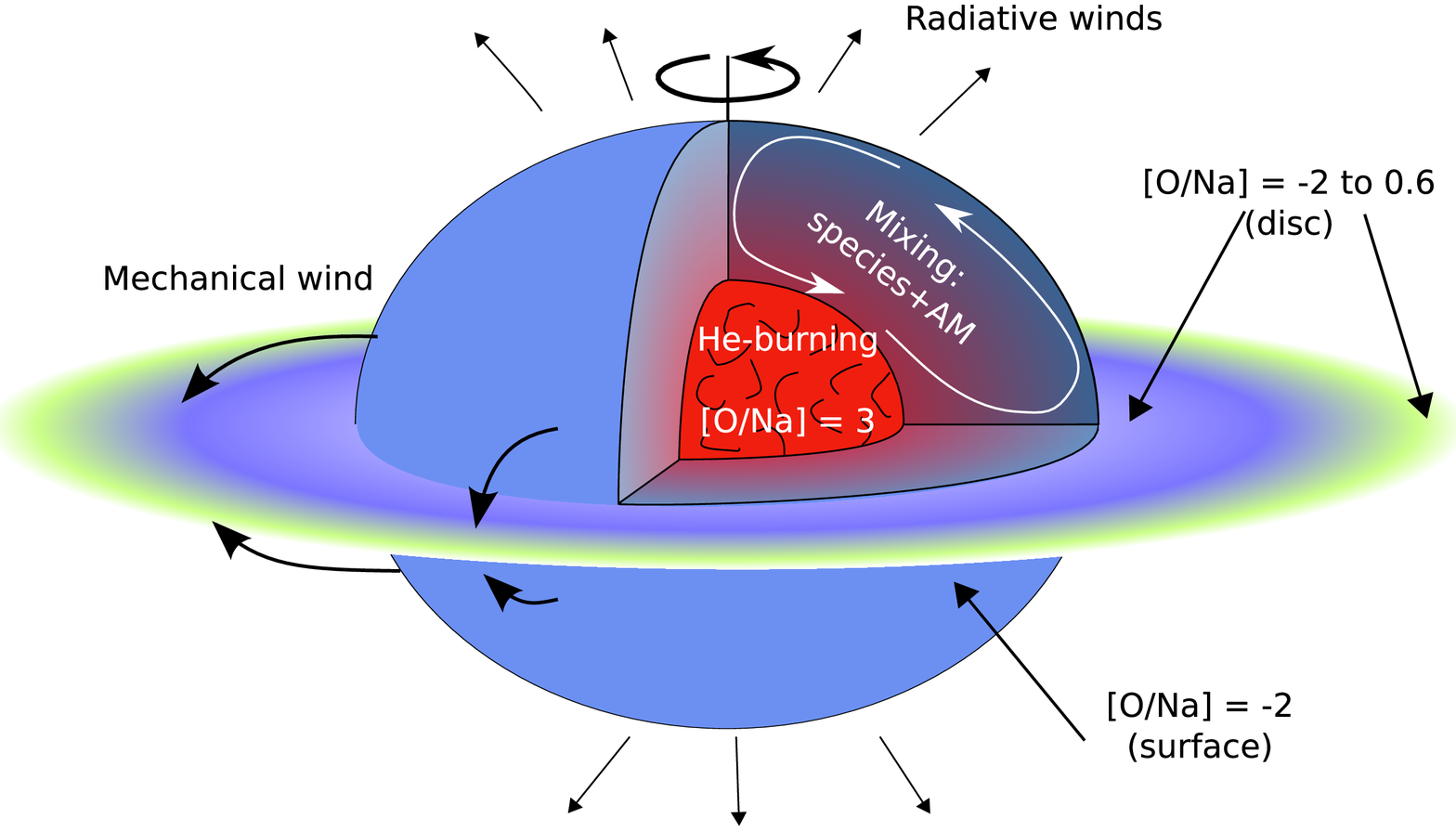}
  \includegraphics[width=0.5\textwidth]{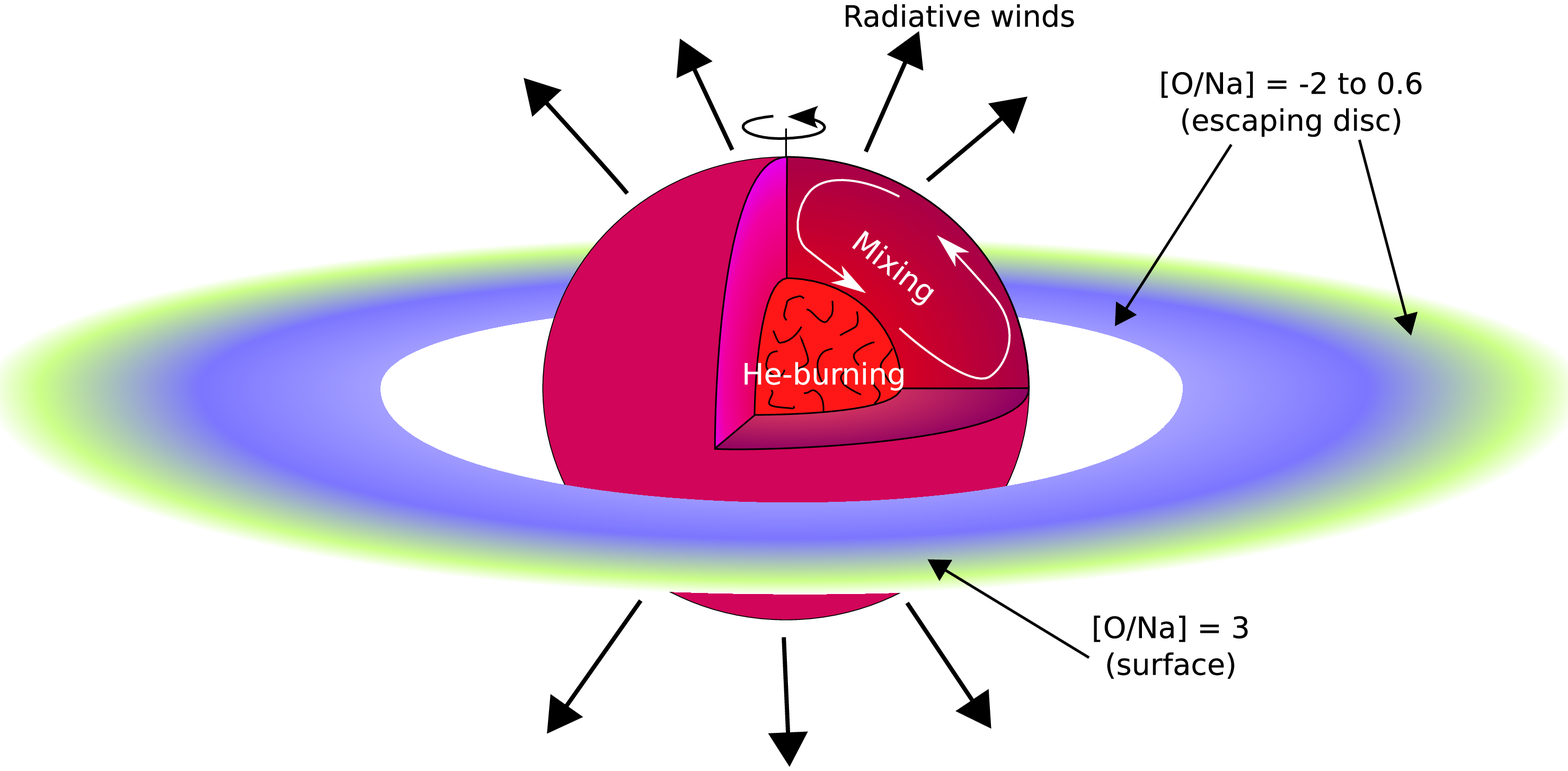}
  \caption{Schematic view of the evolution of fast
    rotating massive stars. The colours reflect the chemical composition of
    the various stellar regions and of the disc (see text for details).
    (top) During the main sequence, a slow outflowing equatorial disc forms
    and dominates matter ejection with respect to radiative winds.
    (middle) At the beginning of central He-burning, the composition of the
    disc material spans the range in [O/Na] observed today in low-mass
    cluster stars. The star has already lost an important fraction of its
    initial mass. (bottom) Due to heavy mass loss, the star moves away
    from critical velocity and does not supply its disc anymore;
    radiatively-driven fast wind takes over before the products of
    He-burning reach the stellar surface.}
  \label{fig:spinstars}
\end{figure}

Any model attempting to explain the chemical properties of globular cluster
stars should give an answer to the following questions: (1) Which type of
stars produced the material enriched in H-burning products?  (2) What is
the physical mechanism responsible for selecting only material bearing the
signatures of H-processing?  (3) Why does this process occur only in
globular clusters?\footnote{Indeed up to now the peculiar chemical patterns
  observed in globular clusters, i.e., the O-Na and Mg-Al anticorrelations,
  have not been found in field stars.}
 
On the basis of considerations on IMF of the potential ``polluting stars'',
\citetalias{PrantzosCharbonnel2006} proposed massive stars (i.e., stars
with an initial mass higher than $\sim$ 10~\Ms{}) as possible ``culprits''.
In \defcitealias{DecressinMeynet2007}{Paper~I}\citet[hereafter
\citetalias{DecressinMeynet2007}]{DecressinMeynet2007} we extended this
idea and developed stellar models sustaining our so-called ``wind of fast
rotating massive stars'' (hereafter WFRMS) scenario that gives satisfactory
answers to the previous questions, and that we recall below.

\section{The wind of fast rotating stars scenario - main guidelines}

In our WFRMS scenario, a first generation of stars forms from proto-cluster
gas already pre-enriched in heavy metals during the halo chemical evolution
(see \citetalias{PrantzosCharbonnel2006} for a more extended
discussion). They have thus the same initial composition as their field
contemporaries; they present in particular the highest [O/Na] ratio we can
observe today in any individual globular cluster. We assume that stars with
masses between 0.1 and 120~\Ms{} form at that stage, with an initial mass
function (IMF) that has to be determined.

Based on observations supporting the view that stars born in dense
environments have faster rotational velocities than stars born in loose
aggregates \citep{Keller2004,StromWolff2005,HuangGies2006,DuftonRyans2006},
we further assume that the first generation massive stars are fast
rotators.  The evolution of such an object is shown in
Fig.~\ref{fig:spinstars}. The colours reflect the chemical composition of
the various stellar regions: Green corresponds to the initial chemical
composition, blue and red are respectively for material loaded in H- and
He-burning products. The typical [O/Na] value in the various stellar
regions is indicated. In the protostars of first generation, [O/Na] is
typically $\sim$ 0.6 (see Sect.~3).
 
Fast rotating massive stars with typical time averaged velocity of
500~km~s$^{-1}$ on the main sequence easily reach the critical
velocity\footnote{The critical velocity is the velocity at the equator such
  that the centrifugal acceleration balances the gravity.} at the
beginning of their evolution and remain near the critical limit during the
rest of the main sequence and part of the central He-burning phase (see
\citetalias{DecressinMeynet2007} and Ekstroem et al. submitted).  As a
consequence, during those phases they lose large amounts of material
through a mechanical wind\footnote{For example, our 60~\Ms{} model
  loses 24.3~\Ms{} in its slow wind.}, which probably leads to the
formation of a slow outflowing Keplerian equatorial disc. This is the kind
of process believed to occur in Be stars \citep{PorterRivinius2003}.
 
The material ejected in the disc has two interesting characteristics: (1)
due to rotational mixing that transports the products of the CNO cycle and
of the Ne-Na and Mg-Al chains from the core to the stellar surface, it
bears the signatures of H-burning and presents abundance patterns similar
to the chemical anomalies observed in the second generation stars as
indicated by the range of [O/Na]\footnote{Early on the main sequence, the
  [O/Na] value in the slow wind is pristine (top panel). As mixing
  proceeds, matter with [O/Na] typical of H-burning (\ie{} $\sim$~-2) is
  ejected in the slow wind (middle panel). In
  \citetalias{DecressinMeynet2007}, we showed that the yields of fast
  rotating massive stars (with an initial mass higher than $\sim$ 20~\Ms{})
  properly explain the O-Na anticorrelation observed today in Galactic
  globular clusters. We note however that in order to reproduce the lowest
  Mg values observed we had to postulate an increase of the
  \el{Mg}{24}($p,\gamma$)\el{Al}{25} nuclear reaction rate with respect to
  the published values.}; (2) it is released into the circumstellar
environment with a very low velocity and thus can easily be retained in
the shallow potential well of the globular cluster. In the following we
shall call it the {\it slow wind}.

The disc-star configuration lasts throughout the main sequence for stars
close to or at critical velocity; for the most massive stars that are at
the $\Omega\Gamma$-limit (\ie{} when the surface luminosity is near the
Eddington limit), it persists in the LBV phase. When the star evolves away
from the critical limit due to heavy mass loss, the radiatively-driven fast
wind takes over and the disc is supplied by the star no longer. This
happens during the central He-burning phase, before the He-burning products
reach the stellar surface and contaminate the slow wind component (bottom
panel of Fig.~\ref{fig:spinstars}). From that moment on, the high-speed
material ejected by the star and by the potential supernova will escape the
globular cluster.
\textit{We proposed that this filtering mechanism, which retains in the
  potential well of the globular cluster only the slow stellar ejecta, is
  the physical mechanism responsible for supplying the required H-processed
  material for forming the stars of the second generation.}

In the vicinity of massive stars, star formation may occur, triggered for
instance by the ionisation front. The large star-to-star spread in light
elements observed today indicates that all the material ejected by the
first generation massive stars did not have the time to be fully mixed
before being recycled in the second generation. There is however clear
evidence from the observed behaviour of both O and Li that the material
ejected in the slow wind is mixed with pristine interstellar matter (ISM;
see
\citealt{KudryashovTutukov1988,DenisenkovDenisenkova1989,DenisenkovDenisenkova1990,LangerHoffman1993,LangerHoffman1995,PrantzosCharbonnel2007}). This
suggests that star formation may have occurred in the vicinity of
individual massive stars from clumps in the slow wind interacting with
pristine gas.  As discussed before, the material ejected in the slow wind
presents various degrees of enrichment in H-burning products. At the
beginning of the disc-star phase, pristine material is ejected giving birth
to stars with a composition similar to that of the first generation. As
time proceeds, the slow wind becomes more and more polluted in H-burning
products (see Fig.~1) and forms stars that are more and more ``anomalous''.
This phase lasts only a few million years
(\eg{} 4.5~Myr for a 60~\Ms{} star).

At this stage, the WFRMS scenario gives promising answers to the questions
raised in Sect.~1, but it does not address the question of the dynamical
evolution of the cluster as a whole.  This is what we explore in
the present paper, as well as other consequences regarding the He content
and the carbon and oxygen isotopic ratios of the second generation stars.
Based on quantitative theoretical yields of our models for fast rotating
massive stars \citepalias{DecressinMeynet2007}, we first address the
question of the IMF of the polluter stars. We use the method developed
by \citetalias{PrantzosCharbonnel2006} which relies on the observed [O/Na]
distribution to derive the amount of polluted material with respect to the
pristine composition.  In Sect.~3 we present the general equations giving the
number ratio of anomalous to normal stars, the number fraction of stars
born with a specific chemical composition, as well as the ratio of the mass
of stellar remnants to stars in the cluster today. The two scenarios that
we have investigated regarding the early evolution of a Galactic globular
cluster are presented in Sect.~4.  Application to the case of NGC~6752 is
discussed in Sect.~5. Section~6 presents our predictions for the helium content
and the isotopic ratios $^{12}$C/$^{13}$C and $^{16}$O/$^{17}$O
respectively. Discussion and conclusions are the subject of Sect.~7.
 
\section{General equations}

Before deriving the equations that will be necessary in this paper, we
recall that only the stars with an initial mass lower or equal to $\sim
0.8$~\Ms{} are still alive today. This corresponds approximately to the
turn-off mass of $\sim 12$~Gyr globular clusters. We will call them the
long-lived (hereafter LL) stars. More massive stars has died previously and
formed dark stellar remnants.

We note also that, according to the nucleosynthetic predictions of
\citetalias{DecressinMeynet2007}, the polluters that are able to provide
the proper material for the second stellar generation consist of stars with
initial masses $\ge 20$~\Ms{}. We assume that all the low-speed
material ejected by those massive stars (at break-up during the main
sequence and during the latter $\Omega\Gamma$-limit phase) is used locally
to form the second stellar generation after having been diluted to various
degrees with pristine ISM. We do not make any assumption regarding the
composition of the stellar ejecta, but rather use our theoretical
predictions for the composition of the slow wind\footnote{In \citetalias{PrantzosCharbonnel2006} the
  estimation of the polluter IMF for NGC~2808 was based on very
  conservative assumptions that maximised the mass of the H-processed
  ejecta released by massive stars. In particular, all the mass outside
  the He-core was assumed to have the proper composition.}.

\subsection{Number ratio of normal to anomalous stars}

The first quantity we need to determine is the number ratio of second
generation (anomalous) stars to the first generation (normal) ones.

If the initial total mass of first generation stars of all masses is
$M_\text{GC}^{1G}$, then the present day mass in first generation
long-lived stars is:
\begin{equation}
  \label{eq:MFG}
  M_\text{LL}^{1G} = M_\text{GC}^{1G} \times \fll{1}\times (1-\elm{1}).
\end{equation}
$\fll{1}$ is the mass fraction locked initially in first generation
long-lived stars; it is given by integration over the IMF in the mass range
0.1--0.8~\Ms.  $\elm{1}$ is the fraction of first generation long-lived
stars that have escaped from the cluster over its history, and reflects the
dynamical properties of the cluster as a whole.  For the sake of simplicity we
assume that this latter parameter is independent of the initial stellar
mass.

Regarding the second generation long-lived stars, we must consider the fact
that they form out of the slow wind of first generation massive stars that
is {\sl diluted} with pristine gas.  The present-day mass locked into those
stars is then:
\begin{equation}
  \label{eq:MSG}
  M_\text{LL}^{2G} = M_\text{GC}^{1G} \times f_\text{SW} \times (1+d) \times
  \fll{2}\times (1-\elm{2}).
\end{equation}
$f_\text{SW}$ is the mass fraction of slow winds produced by massive stars,
which is given by theoretical stellar models.  $d$ is the parameter
reflecting global dilution with pristine ISM; it can be inferred from
detailed analysis of the observed abundance patterns,
as discussed in Sect.~5.5.  $\fll{2}$ and $\elm{2}$ are respectively the
initial mass fraction of the second generation long-lived stars and the
fraction of these objects that have escaped the cluster (as for $\elm{1}$
we assume that $\elm{2}$ is independent of the initial stellar mass).

Therefore, the number ratio between these two populations of long-lived
stars is:
\begin{equation}
  \label{eq:number}  
  \frac{n_\text{LL}^\text{2G}}{n_\text{LL}^\text{1G}} =
  \frac{\langle M_\text{LL}^\text{1G}\rangle}{\langle M_\text{LL}^\text{2G}
    \rangle} \times f_\text{SW} \times (1+d)
  \times \frac{\fll{2}}{\fll{1}} \times
  \frac{1- \elm{2}}{1-\elm{1}},
\end{equation}
where $\langle M_\text{LL}^\text{1G} \rangle$ and $\langle
M_\text{LL}^\text{2G} \rangle$ are the average masses of the long-lived
stars of first and second generation respectively. The left-hand part of
this equation can be inferred from observational studies, following the
method proposed by \citetalias{PrantzosCharbonnel2006} and based on the
observed [O/Na] distribution; this point is discussed further in Sect.~5.6.
The number of escaping stars is a free parameter that is related to the
cluster dynamics.

The stellar IMF of both generations influences the quantities
$f_\text{SW}$, $\fll{1}$, and $\fll{2}$, as well as the ratio $\langle
M_\text{LL}^\text{1G}\rangle/\langle M_\text{LL}^\text{2G} \rangle$. We
follow \citetalias{PrantzosCharbonnel2006} and assume that the IMF of the
globular cluster is composite. For stars of first generation more
massive than 0.8~\Ms{}, the IMF is approximated by a power-law of the form:
\begin{equation}
  \label{eq:phih}
  \Phi(M) =  \frac{\text{d}N}{\text{d} M} \propto M^{-(1+x)},
\end{equation}
with $x$ the slope that has to be determined ($x=1.35$ being the \citealt{Salpeter1955} value).  On the other hand we consider that the present-day log-normal
distribution derived by \citet{ParesceDeMarchi2000} reflects the IMF of
long-lived stars (0.1-0.8~\Ms{}) of both first and second generations:
\begin{equation}
  \label{eq:phil}
  \ln \Phi(M) 
  \propto A - \left( \frac{\log(M/M_\text{C})}{2\sigma}\right)^2,
\end{equation}
where $M_\text{C}=0.33\pm0.03$ is the peak-mass of the log-normal
distribution, $\sigma=0.34\pm0.04$ a standard deviation, and A a
normalisation constant.
We normalise all IMF to unity:
\begin{equation}
  \label{eq:norm}
  \int_{0.1}^{120} M \Phi(M) \text{d}M = 1.
\end{equation}
Since we presume the same IMF for the first and second generation
long-lived stars,
one has that $\langle M_\text{LL}^\text{1G}\rangle=\langle
M_\text{LL}^\text{2G}\rangle$.

\subsection{Chemical composition of second generation stars}

The next step consists of the derivation of the equations required to
determine the chemical abundance distribution in the matter out of which
the second generation forms. This distribution will reflect directly in the
star-to-star abundance variations at their birth. It will also reflect the
distribution in the star-to-star abundance variations that we observe today
in stars that have not undergone any change of their surface abundances by
whatever internal mixing process, {\it i.e.} typically for stars having not
yet evolved through their first dredge-up\footnote{These stars may however
  have suffered from some depletion of their lithium surface abundance.}.
We recall that this matter consists of the slow winds of individual stars
of first generation diluted locally and to various degrees with pristine
ISM. As discussed in Sect.~2, the degree of enrichment of the slow wind by
the H-burning products is likely to be different at the beginning and the
end of the evolution of the massive polluter. Let us call $t$ the age of
the massive star polluter, then
\begin{equation}
  \label{eq:dilution}
  X_i^\text{2G} (a_t) = (1-a_t) X_i^\text{SW}(t) + a_t X_i^\text{init}
\end{equation}
with $X_i^\text{init}$ and $X_i^\text{SW}$ being the mass fraction of the
element $i$ respectively in the pristine ISM and in the slow wind of the
polluter. $X_i^\text{2G}$ corresponds to the same quantity after dilution
of the massive star ejecta with pristine material, $a_t$ being the local
dilution parameter (see more detail in Sect.~5.5).  Let us call $\dot
m_\text{SW}(t, M)$ the rate of ejection of mass in the slow wind at time
$t$ by a star of given initial mass $M$. The mass ejected during the time
$\diff t$ is $\dot m_\text{SW}(t, M)\diff t$. After dilution, the matter
available to form the second generation stars from this mass is $\dot
m_\text{SW}(t)\diff t/(1-a_t)$. Integrating over the duration of the ``slow
wind'' phase, $\Delta t$ and over the IMF, one finds an expression for the
matter available to form the second generation stars,
\begin{equation}
  \label{eq:M2Gbis}
  M_\text{GC}^{2G} = M_\text{GC}^{1G}\int_{20}^{120} \left( \int_{\Delta t}
    \dot m_\text{SW}(t,M) \frac1{1-a_t} \diff t \right) \Phi(M)
  \diff M.
\end{equation}
In this case, the present-day mass locked into these stars is then
\begin{equation}
  \label{eq:M2Gter}
  M_{LL}^{2G} = M_\text{GC}^{2G}\times
  \fll{2}\times (1-\elm{2})
\end{equation}
By identification between Eqs.~(\ref{eq:M2Gter}) and (\ref{eq:MSG}) we can
obtain a relation between the global dilution parameter $d$ and the local
one $a_t$:
\begin{equation}
  \label{eq:globloc}
  d = \frac{1}{f_\text{SW}} \int_{20}^{120} \int_{\Delta t}  \dot m_\text{SW}(t,M)
  \frac{a_t}{1-a_t} \diff t \Phi(M) \diff M.
\end{equation}
In order to obtain the mass available to form second generation stars with
a given abundance of the element $i$ equal to $X_i^\text{2G}$, the integral
over time in Eq~(\ref{eq:M2Gbis}) has to be restrained to that
portion of the slow wind phase, during which the wind has a given abundance
of the element $i$ equal to $X_i^\text{SW}$, where $X_i^\text{SW}$ is the
abundance such that, after dilution, the material made of slow wind and
ISM will have an abundance equal to $X_i^\text{2G}$. If we 
call $\Delta t_i$ this duration, then, the mass available to form second
generation stars with a given abundance of the element $i$ equal to
$X_i^\text{2G}$ is given by
\begin{equation}
  \label{eq:M2GXi}
  M_{GC}^{2G}(X_i^\text{2G}) = M_\text{GC}^{1G}\int_{20}^{120} \left( \int_{\Delta t_i}
    \dot m_\text{SW}(t,M) \frac1{1-a_t} \diff t \right) \Phi(M)
  \diff M,
\end{equation}
and the present-day mass locked into those stars is then
\begin{equation}
  \label{eq:M2GXibis}
  M_{LL}^{2G}(X_i^\text{2G}) = M_\text{GC}^{2G}(X_i^\text{2G})\times
  \fll{2}\times (1-\elm{2}).
\end{equation}
Normalised distribution is obtained by dividing Eq.~(\ref{eq:M2GXibis}) by
the total number, $M_\text{LL}$, of first and second generation stars given
respectively by Eq.~(\ref{eq:MFG}) and Eq.~(\ref{eq:MSG}):
\begin{equation}
  \label{eq:normdistr}
  \frac{M_\text{LL}^{2G}(X_i^\text{2G})}{M_\text{LL}} = \frac{M_\text{GC}^{2G}(X_i^\text{2G}) \times
    \fll{2}\times (1-\elm{2})}{
    f_\text{LL}^{1G}(1-e_\text{LL}^\text{1G}) + f_\text{SW}(1+d)
    f_\text{LL}^\text{2G}(1-e_\text{LL}^\text{2G})}.
\end{equation}
Note that the above mass ratio is also equal to a number ratio provided the
mean mass of the low mass stars in the first and second generation are the
same, which is the hypothesis we have made here.

We will use the theoretical predictions of our models of fast rotating
massive stars published in \citetalias{DecressinMeynet2007}. In
particular, we have that the stars with initial masses of 20, 40, 60, and
120~\Ms{} expel respectively 1.7, 12.5, 24.3, and 48.0~\Ms{} in total in
their slow wind. In order to convolve by the polluter IMF we interpolate
between those values for polluter stars with different initial masses.

\subsection{Mass of stellar remnants}

Another important quantity is the contribution of stellar remnants (i.e.,
white dwarfs, neutron stars or stellar black holes) to the total cluster
mass.  Indeed first and second generation stars more massive than the
so-called long-lived stars (i.e., with initial masses higher than
0.8~\Ms{}) already died and produced dark remnants\footnote{As
  mentioned in Sect.~3.2, we will assume later that only long-lived stars
  formed in the second generation. However we derive here the most general
  equations.}. We call $f_\text{rem}^{1G}$ and $f_\text{rem}^\text{2G}$ the
mass fraction to the total stellar mass of respectively first and second
generation remnants:
\begin{equation}
  \label{eq:fracrem}
  f_\text{rem} = \int_{0.8}^{120} R(M) \Phi(M) \diff M.
\end{equation}
We follow \citetalias{PrantzosCharbonnel2006} for the relations between the
initial mass of a star and the mass of its remnant:
\begin{equation}
  \label{eq:remmass}
  R(M) = \left\{
    \begin{array}{ll}
      0.446+0.106M & \text{for } M<10\Ms\\
      1.5          & \text{for } 10<M<25\Ms \\
      3            & \text{for } M>25\Ms\\
    \end{array}
  \right.
\end{equation}
The total mass of the remnants in the cluster today is given by:
\begin{equation}
  \label{eq:remnants}
  M_\text{rem} = M_\text{GC}^{1G} f_\text{rem}^{1G}(1-e_\text{rem}^\text{1G}) +
  M_\text{GC}^{1G} f_\text{SW} (1+d) f_\text{rem}^\text{2G}(1-e_\text{rem}^\text{2G}),
\end{equation}
with $e_\text{rem}^\text{1G}$ and $e_\text{rem}^\text{2G}$ the fraction of
remnants lost by the cluster during its evolution (these quantities take
into account the loss of both remnant progenitors and remnants).

Therefore the mass ratio between remnants and long-lived stars today is
obtained by combining Eq.~(\ref{eq:MFG}), (\ref{eq:MSG}), and
(\ref{eq:remnants}), i.e.:
\begin{equation}
  \label{eq:remtolm}
  \frac{M_\text{rem}}{M_\text{LL}} = \frac{
    f_\text{rem}^{1G}(1-e_\text{rem}^\text{1G}) + f_\text{SW}(1+d)
    f_\text{rem}^\text{2G}(1-e_\text{rem}^\text{2G})} 
  {f_\text{LL}^{1G}(1-e_\text{LL}^\text{1G}) + f_\text{SW}(1+d)
    f_\text{LL}^\text{2G}(1-e_\text{LL}^\text{2G})}.
\end{equation}

\begin{figure}[tbp]
  \includegraphics[width=0.48\textwidth]{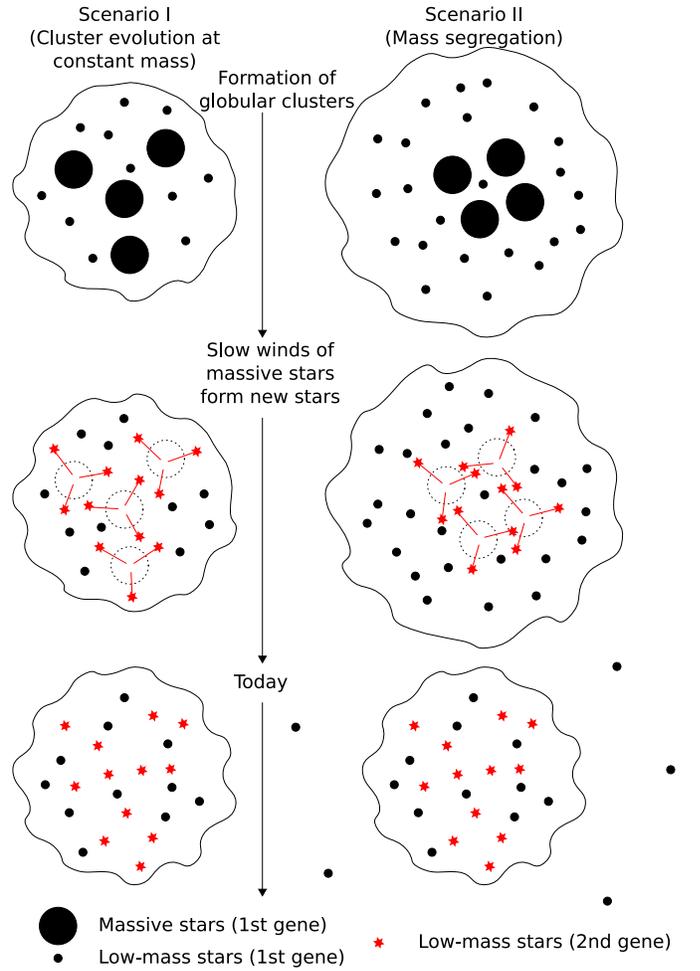}
  \caption{Schematic illustration of the scenarios discussed in
    Sect.~4. Black circles and red stars stand for first and second
    generation stars respectively. In the middle panels, the dashed symbols
    represent fast evolving massive stars giving birth to second generation
    stars. In both cases, only long-lived low-mass stars form in the second
    generation. In Scenario I (left panels) all the stars remain in the
    cluster and the results depend on the chosen IMF. In Scenario II (right
    panels), we assume mass segregation and pollution occurs only at the
    center of the cluster; the first generation long-lived stars that
    remained in the outer part are assumed to be lost over the whole
    dynamical evolution.}
  \label{fig:shemas}
\end{figure}

\section{Our two scenarios for the dynamical evolution of the cluster}

As some of the parameters used in the previous equations are poorly known
and are related to general properties of globular clusters (i.e., their IMF
and their dynamical evolution) that are poorly constrained, we explore
two simplifying scenarios. For the sake of simplicity, we make the following
assumptions in both cases. We suppose that the first stellar generation
contains stars with initial masses between 0.1 and 120~\Ms{}.
Regarding the second generation, we assume that it consisted only of
low-mass long-lived stars (i.e., with an initial mass between 0.1 and
0.8~\Ms{}); this latter assumption will obviously minimise the
constraint on the polluter IMF.

\subsection{Scenario I: no loss of stars and of stellar remnants}
 
The first scenario we consider (left panels in Fig.~\ref{fig:shemas})
assumes that the cluster as a whole has retained all its stars and stellar
remnants. In other words, the cluster stellar component underwent no
evaporation of stars due for instance to tidal stripping.
In this case one has thus
\begin{equation}
  \label{eq:scen1evap}
  \elm{1} = \elm{2} = \erem{1} = \erem{2} = 0.
\end{equation}

Since we assume that only low-mass stars are produced in the second
generation, we have
\begin{equation}
  \label{eq:scen1lm}
  \fll{2} = 1 \text{ and } \frem{2} = 0.
\end{equation}
We also assume that the slope of the IMF is the same as far as the first
and second generation long-lived stars are concerned.  As a result of these
simplifications, the general equations derived in Sect.~3 become
\begin{equation}
  \label{eq:s1a}
  \frac{n_\text{LL}^\text{2G}}{n_\text{LL}^\text{1G}} = \frac{ f_\text{SW}
    (1+d)} {\fll{1}},
\end{equation}
for the number ratio between normal and anomalous long-lived stars (see
Eq.~\ref{eq:number}),
\begin{equation}
  \label{eq:s1b}
  \frac{M_\text{LL}^{2G}(X_i^\text{2G})}{M_\text{LL}} =\frac{
    \int_{20}^{120} \left( \int_{\Delta t_i}
      \dot m_\text{SW}(t,M) \frac1{1-a_t} \diff t \right) \Phi(M)
    \diff M}{
    f_\text{LL}^{1G}+ f_\text{SW}(1+d)},
\end{equation}
for the number fraction of stars of second generation with a given
composition (see Eq.~\ref{eq:normdistr}), and
\begin{equation}
  \label{eq:s1c} 
  \frac{M_\text{rem}}{M_\text{LL}} = \frac{ f_\text{rem}^{1G} }
  {f_\text{LL}^{1G} + f_\text{SW}(1+d)}
\end{equation}
for the fraction of the total mass in stellar remnants (see
Eq.~\ref{eq:remtolm}).

The free parameters in Scenario I are the slope of the IMF of the polluters
and the dilution factor.  It should be noted that the Scenario I with no
evaporation of stars is equivalent to that derived in the case where stars
of both first and second generation leave the cluster in a similar way
({\it i.e.} where $\elm{1} = \elm{2} = \erem{1} = \erem{2}$ but are not
necessarily equal to 0 as in Eq.~\ref{eq:scen1evap}).

\subsection{Scenario II: mass segregation and evaporation of stars}

In Scenario II, we assume (see right panels in Fig.~\ref{fig:shemas}) that
the massive polluters of first generation were born in the center of the
cluster or have migrated very rapidly towards this region. In this case,
the second generation stars are created only in the central region while
the external part of the cluster hosts only first generation long-lived
low-mass stars. This strong radial distribution is assumed to stay in place
until the moment where supernovae sweep away the residual intra-cluster gas
and thus strongly modify the cluster potential well. As a consequence,
stars in the external part of the cluster do not remain bound and are
ejected into the galactic halo.

In this case, we impose a \citeauthor{Salpeter1955} IMF for the
polluters\footnote{It should be noted that this IMF is not valid for the
  whole mass range of polluters. However our findings would not change
  significantly with a more realistic IMF as that of \citet{KroupaTout1993}},
and search for the amount of first generation long-lived stars that need to be
lost in order to fit the observational constraints.
As a result of our hypothesis, one has $\fll{2} = 1$, $\elm{2} = 0$, while
$\elm{1}$ remains a free parameter. The general equations become (see
Eq.~\ref{eq:number}, \ref{eq:normdistr} and \ref{eq:remtolm}):
\begin{equation}
  \label{eq:s2a}
  \frac{n_\text{LL}^\text{2G}}{n_\text{LL}^\text{1G}} =
  \frac{f_\text{SW} (1+d)} {\fll{1} \left( 1-\elm{1} \right)},
\end{equation}
\begin{equation}
  \label{eq:s2b}
  \frac{M_\text{LL}^{2G}(X_i^\text{2G})}{M_\text{LL}} = \frac{
    \int_{20}^{120} \int_{\Delta t_i}
    \dot m_\text{SW}(t,M) \frac1{1-a_t} \diff t \Phi(M) \diff M}{
    f_\text{LL}^{1G}(1-e_\text{LL}^\text{1G}) + f_\text{SW}(1+d)},
\end{equation}
and
\begin{equation}
  \label{eq:s2c}
  \frac{M_\text{rem}}{M_\text{LL}} = \frac{
    f_\text{rem}^{1G}(1-e_\text{rem}^\text{1G}) } 
  {f_\text{LL}^{1G}(1-e_\text{LL}^\text{1G})+ f_\text{SW}(1+d)}.
\end{equation}
The free parameters in Scenario II are the number fractions of stars lost
by the clusters and the dilution factor.

\section{The case of NGC~6752}

Armed with these equations, we will first focus on the globular cluster
\object{NGC~6752} which has about the same metallicity ([Fe/H]~$\sim$~-1.56,
\citealt{CarrettaBragaglia2007}) and initial abundance pattern as the
stellar models presented in \citetalias{DecressinMeynet2007}.  Our
theoretical nucleosynthetic predictions can thus be directly
applied. Additionally, NGC~6752 is the Galactic globular cluster with the
largest set of abundance data for numerous chemical elements (see
references in \citetalias{DecressinMeynet2007}).

\subsection{Observed ratio of anomalous to normal stars in NGC 6752}

\begin{figure}[htpb]
  \includegraphics[width=0.5\textwidth]{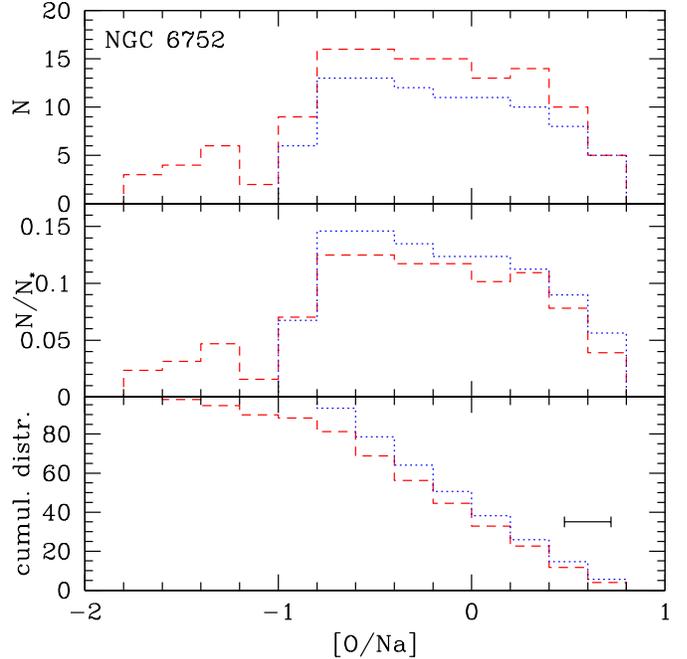}
  \caption{ONa distribution function in NGC~6752 based on observations of
    red giant branch stars from \citet{CarrettaBragaglia2007}. (top)
    Distribution function in a number issue taken from
    \citet{CarrettaBragaglia2007}; (middle) normalised distribution
    function, $N_*$ is the total number of observed stars; (bottom) inverse
    cumulative distribution function for the same stars. Average [O/Na]
    error is also indicated. Dotted lines refer to stars for which O and Na
    abundances have been determined. A sample of stars have determinations
    of the Na abundance but not of O. In this case the abundance of O has
    been inferred from the Na abundance using the global anticorrelation
    between these two elements. Inclusion of this sample of stars gives
    the distributions shown by the dashed lines.}
  \label{fig:ona6752}
\end{figure}

To determine the ratio between the first generation normal and second
generation anomalous long-lived stars in NGC~6752, we use the method
described in \citetalias{PrantzosCharbonnel2006} and base our estimation on
the [O/Na] distribution function derived from the observations by
\citet[see Fig.~\ref{fig:ona6752}]{CarrettaBragaglia2007}. Typical
dispersion errors on O and Na abundances in the sample of 120 stars are
respectively $\sigma_\text{O}=0.101$~dex and
$\sigma_\text{Na}=0.060$~dex. The total dispersion on the [O/Na] ratio is
thus $\sigma_\text{ONa} = \sqrt{\sigma_\text{O}^2+\sigma_\text{Na}^2} =
0.117$, so that the first generation stars span a range of
$\Delta[\text{O/Na}]= 2 \sigma_\text{ONa} = 0.234$. As this value is higher
than the bin width of 0.2 used for the distribution functions displayed in
Fig.~\ref{fig:ona6752}, first generation stars can be found in the two bins
with the highest [O/Na] ratio. They represent between 12\% and 14\% of the
total number of long-lived stars. In the following, we round up to 85/15
the ratio between second and first generation long-lived stars.

The percentage of first generation long-lived stars we find in NGC~6752 is
lower than the values found in previous studies.
\citetalias{PrantzosCharbonnel2006} find indeed 30\% of normal stars in
NGC~2808 while \citet{CarrettaGratton2005} infer that less than 1/3 of the
stars are normal on the basis of CNO abundances in NGC~6397, NGC~6752, and
47~Tuc. This latter study does not however use very precise statistics as
it mixes several clusters. This is discussed further in
Sect.~\ref{sec:discussion}.

\subsection{IMF slope for the polluter stars in Scenario I}
\label{sec:imf-slope}

\begin{figure}[tb]
  \includegraphics[width=0.5\textwidth]{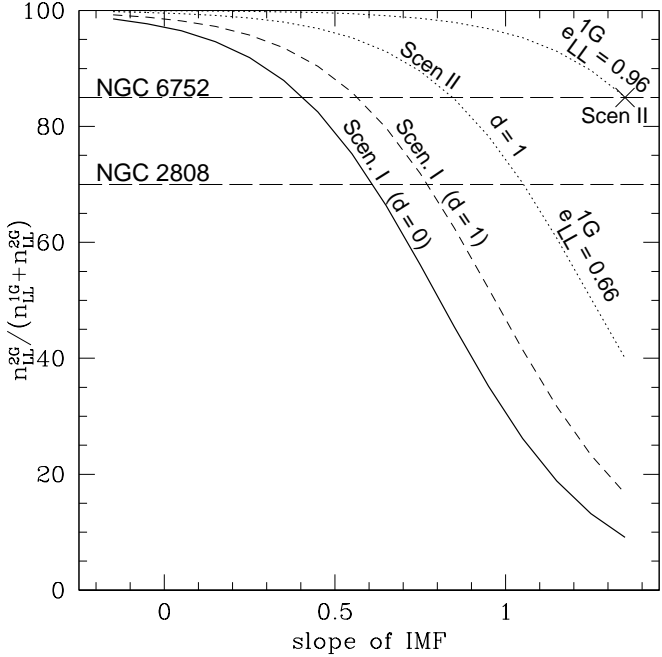}
  \caption{Percentage of second generation stars in the total long-lived
    stars observed today as a function of the IMF slope of the first
    generation polluters. The solid line refers to long-lived stars born
    out of the slow winds of massive stars only, while the dashed line
    corresponds to the case where dilution of the ejecta with pristine ISM
    is taken into account with a global dilution factor $d=1$. Dotted lines
    show the cases where $d=1$, $\elm{1} = 0.66$ and 0.96. The cross
    indicates the point corresponding to a Salpeter IMF and to 85\% of
    second generation long-lived stars. Horizontal lines indicate the
    constraints coming from the observed [O/Na] distribution for two
    well-studied Galactic globular clusters.}
  \label{fig:slope}
\end{figure}

In Scenario~I, the number fraction of second to first generation stars (see
Eq.~\ref{eq:s1a}) depends on the slope of the IMF of the polluters and on
the global dilution factor $d$. Figure~\ref{fig:slope} displays the
fraction of second generation stars as a function of the IMF slope of the
polluters and for dilution factors $d=0$ and 1.
  
A Salpeter IMF with no differential evaporation leads to a small number of
second generation stars that total up 10-17\% of the cluster long-lived
stars depending on the dilution factor (0 or 1). In order to have 85\% of
second generation long-lived stars in the framework of Scenario I, the IMF
slope must be around 0.4 if $d=0$ (full line). This value is a lower limit
as it does not take into account dilution of the slow wind ejecta with some
pristine gas. For a global dilution parameter $d = 1$, the IMF slope raises
up to 0.55 (see the dashed line). Estimating in the same way the IMF slope
of the polluters for the cluster NGC 2808, which has a higher fraction of
normal first generation stars, results in steeper IMF (0.60 and 0.76 for
respectively $d=0$ and 1), although still quite flat with respect to the
Salpeter one. Thus, as already discussed in
\citetalias{PrantzosCharbonnel2006}, Scenario~I implies relatively flat
IMFs for the polluters.

\subsection{Fraction of stars lost by the cluster in Scenario II}

We consider now Scenario II where we use a \citeauthor{Salpeter1955} IMF
for first generation massive stars. A dilution factor $d=1$ is used. In
order to explain the high number of second generation anomalous stars
observed today, a strong loss of first generation long-lived stars from the
cluster is required. More precisely, we derive $\elm{1} \simeq 0.958$ (see
the corresponding dotted curve in Fig.~\ref{fig:slope}).This means that, in
the present day globular cluster, only $\sim 4.2$\% of the first generation
long-lived stars are still present. Note that this is a lower limit;
indeed, if some second generation stars also escaped in the halo, then even
more first generation stars must have been lost by the cluster in order to
reproduce the high number of second generation anomalous stars observed
today in NGC 6752. It should be noted that the unpolluted pop. II stars
found in the halo would comes largely from lower-mass clusters that
dissolved after residual gas expulsion, as pointed out by
\citet{KroupaBoily2002}.

An intermediate case between Scenario I and II is shown in
Fig.~\ref{fig:slope} (dotted line labelled with $\elm{1} = 0.66$). There we
assume that one third of first generation long-lived stars remain in the
cluster. In this third case, an intermediate IMF slope of 0.85 is
needed in order to reproduce the observed ratio of anomalous to normal
stars in NGC 6752.

\subsection{Amount of stellar residues}

\begin{figure}[tbp]
  \includegraphics[width=0.5\textwidth]{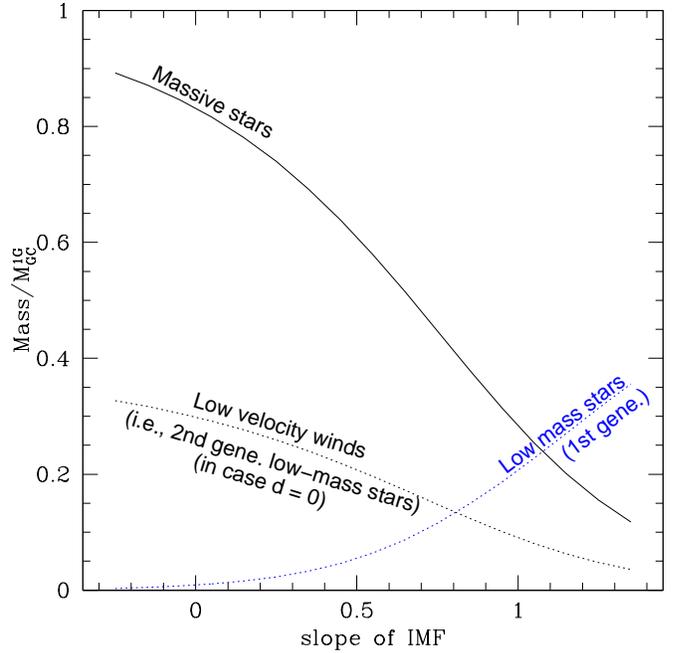}
  \caption{Mass fractions of different stellar populations at birth
    normalised to the mass of the first generation stars, as a function of
    the slope of the polluter IMF. Full line refers to the massive stars
    (20--120~\Ms). Dotted lines indicate the long-lived stars
    (0.1--0.8~\Ms) of first and second generations (see the labels of the
    curves). The mass fraction of second generation stars corresponds to
    the case without dilution.}
  \label{fig:imfpop}
\end{figure}

\begin{figure}[tbp]
  \includegraphics[width=0.5\textwidth]{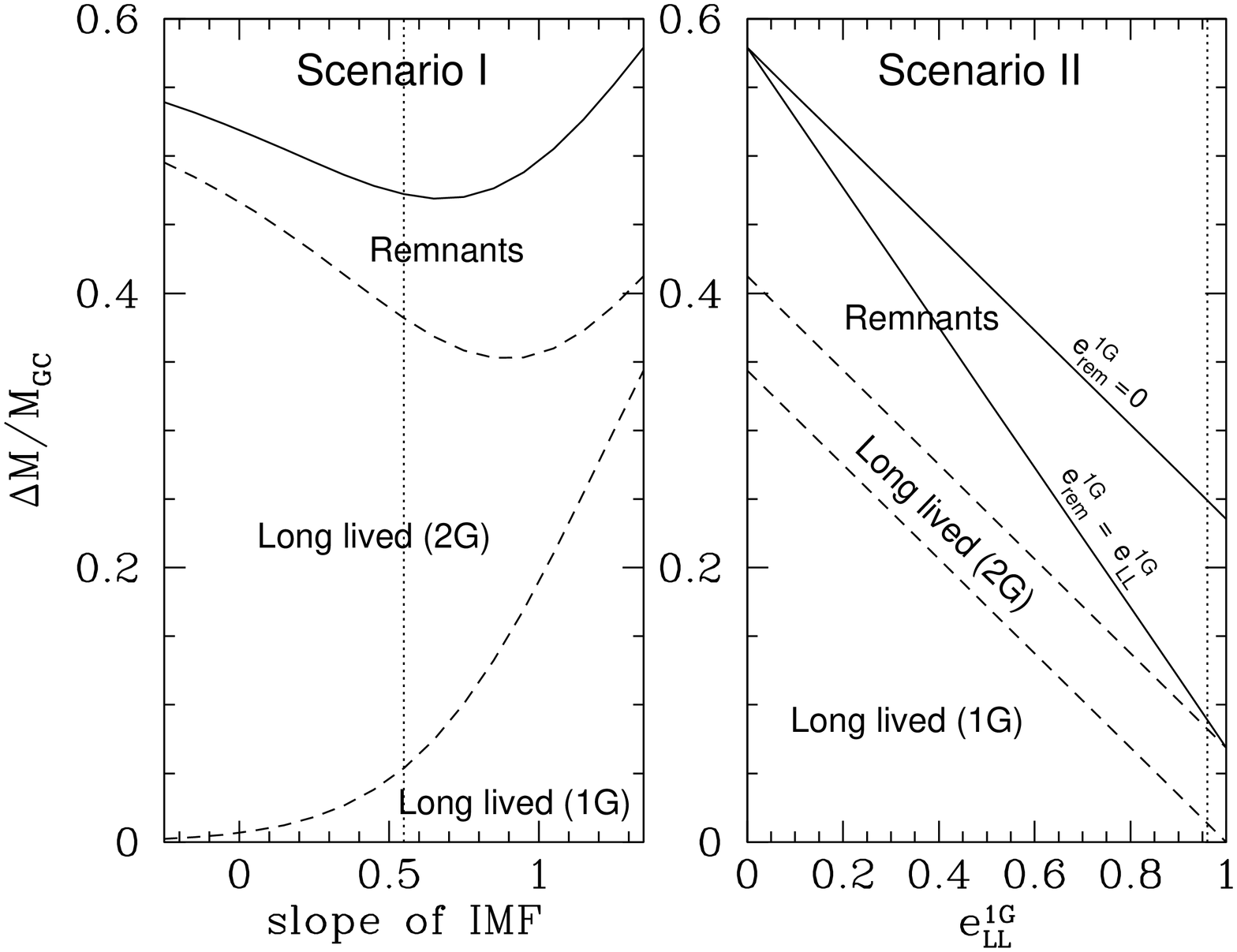}
  \caption{Variation of the present day total mass of the cluster
    (continuous line) as a function of the slope of the IMF (Scenario I,
    left panel) and as a function of the fraction of stars of the first
    generation which have been lost by the cluster, $e_\text{LL}^\text{1G}$
    (Scenario~II, right panel). The vertical extension of each of the three
    labeled zones gives the mass fraction of the cluster locked into long
    lived stars of first and second generation stars and in remnants. The
    line labeled as $e_\text{rem}^\text{1G}=0$ and
    $e_\text{rem}^\text{1G}=e_\text{LL}^\text{1G}$ indicate respectively
    the case where no remnant are lost by the clusters and where they have
    been lost at the same rate as long-lived stars.  The vertical dotted
    line shows the ratio of second to first generation stars is equal to
    the observed value of 85/15.}
  \label{fig:IMFe1and2}
\end{figure}

Figure.~\ref{fig:imfpop} displays the mass fraction of different stellar
populations at birth ({\it i.e.} before any evaporation of stars) as a
function of the IMF slope for the polluters. Normalisation in made with the
total mass of gas used to form stars (\ie{} the gas used to form the first
stellar generation added to the one used for the dilution process at the
birth of anomalous stars).  As the IMF slope decreases, massive stars
initially dominate the cluster mass; for an IMF of $\sim$ 0.55, the matter
ejected in the slow winds represents $\sim 20$\% of the total mass of all
first generation stars and more than three times the mass locked into long
lived first generation stars.

The left panel of Fig.~\ref{fig:IMFe1and2} shows how the total stellar mass
of the present day cluster varies as a function of the polluter IMF in the
frame of Scenario I (continuous line). The mass is normalised to the mass
of gas used to form stars.  The following remarks can be made:
\begin{itemize}
\item We see that the stellar component in the present day cluster
  represents about half of the total mass of gas used to form stars (more
  precisely between $\sim$47 and 57\% depending on the slope of the IMF for
  the polluters). The rest of the mass consists of material ejected by the
  stars under the form of fast ejecta escaping the potential well of the
  cluster (fast winds of massive stars, supernova ejecta, winds of red
  giants and asymptotic giant branch stars).
\item The way the total mass varies as a function of the slope of the IMF
  does appear at first sight quite surprising. Why do we obtain a
  non-monotonic behaviour?  We note that if there had 
  been only a unique star formation episode, the present day mass cluster
  would be lower than the mass contained in the gas used to form stars. As
  explained above, the cluster loses fast stellar ejecta. Since their
  proportion increases when the slope of the IMF is decreased, smaller
  masses of the present day clusters are obtained with flatter IMF.  In our
  case, the situation is not so straightforward, since part of the massive
  star ejecta are actually released under the form of slow winds which are
  retained by the clusters and are used to form the second generation
  stars. Therefore, the increase of the number of massive stars ({\it i.e.}
  a decrease of the slope of the IMF) may produce either a decrease or an
  increase of the present day mass cluster, depending on how the mass lost
  by the fast ejecta is compensated by the second generation stars formed
  from the slow ejecta and from some amount of ISM. From
  Fig.~\ref{fig:IMFe1and2}, we see that for an IMF slope inferior to $\sim$
  0.75, the mass locked in second generation stars is greater than the mass
  lost through fast ejecta, while for slopes of the IMF superior to 0.75,
  the opposite situation occurs.
\item The population of remnants never exceeds those of long lived
  stars. At most, it can contain about one third of the present day mass in
  the case of a Salpeter IMF. The vertical line at an abscissa of 0.55
  corresponds to a ratio of anomalous (second generation) to normal (first
  generation) star equal to 85/15.
\end{itemize}

The right panel of Fig.~\ref{fig:IMFe1and2} shows how the present day mass
of the cluster varies as a function of the fraction of stars of the first
generation which escape the cluster. Since an IMF slope of 1.35 is chosen
in that case, we obtain for $e_\text{LL}^\text{1G}=0$ the same results as
in Scenario I at the corresponding IMF. As already obtained in
Fig.~\ref{fig:slope}, we see that to obtain a ratio of anomalous (second
generation) to normal (first generation) star equal to 85/15 (see the
vertical line with an abscissa equal to 0.958), a very large fraction of
first generation long lived stars should have been lost by the cluster.
For this value of $e_\text{LL}^\text{1G}$, in the case that no remnant has
escaped ($e_\text{rem}^\text{1G}=0$), the mass of the cluster would be
dominated by the stellar residues. Only when a large fraction of the
residues also escape the cluster
($e_\text{rem}^\text{1G}=e_\text{LL}^\text{1G}$), would its mass be
dominated by the stars.  In that case the present day mass of the cluster
represents less than 10\% of the mass of the gas used to form stars.

\subsection{Amount of dilution}
\label{sec:dilution-process}

\begin{figure}[htbp]
  \includegraphics[width=0.5\textwidth]{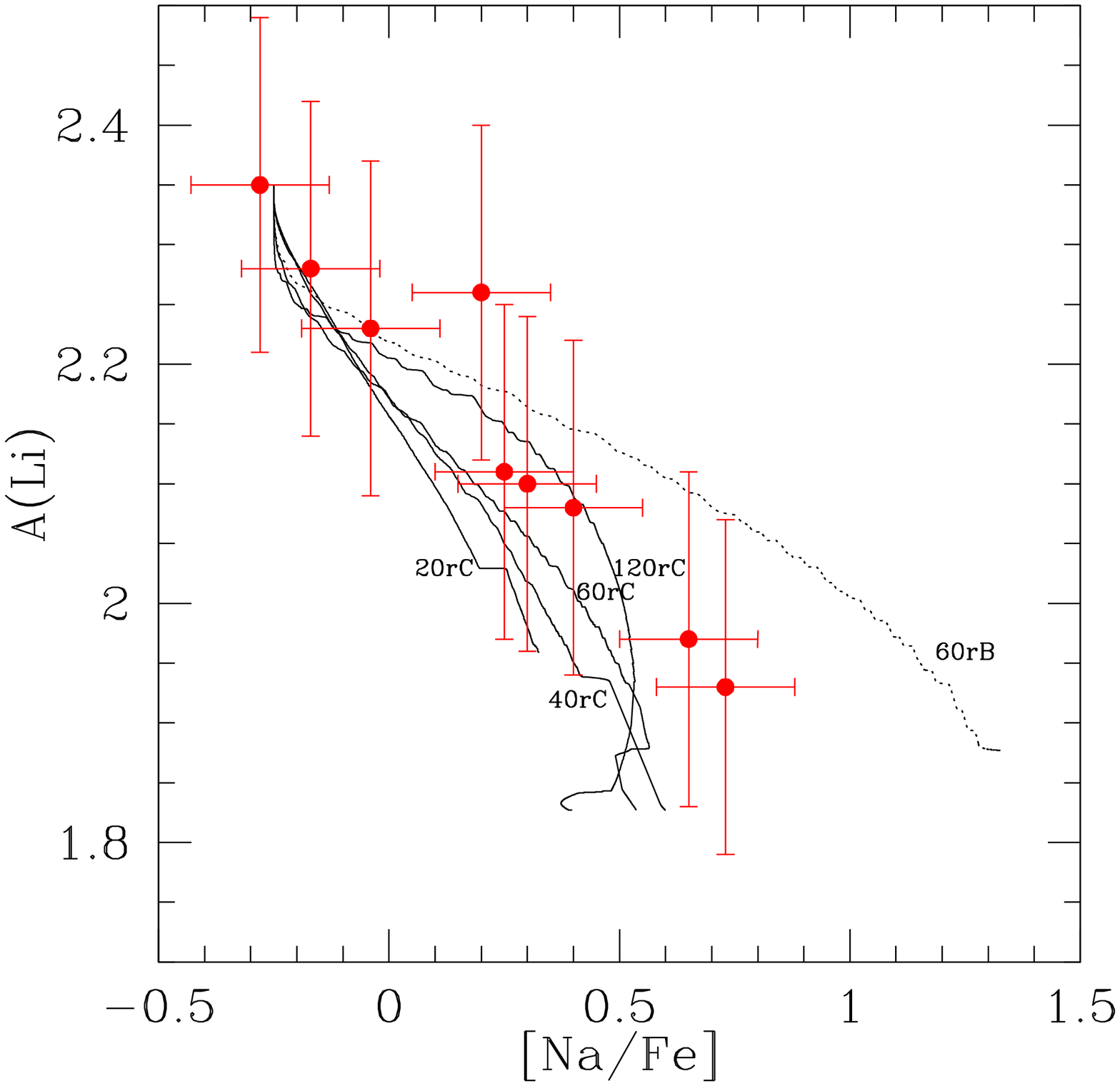}
  \caption{Li depletion vs [Na/Fe] value. Full circles indicate the
    observations of turnoff stars by \citet{PasquiniBonifacio2005} in
    NGC~6752. The theoretical predictions for the Li abundance (full and
    dotted lines) have been decreased by 0.3~dex to take into account
    in-situ Li depletion during the main sequence of low-mass stars (see
    \citetalias{PrantzosCharbonnel2006} and \citealt{CharbonnelPrimas2005}
    for more details). Full lines refer to Li and Na in the
    ejecta of individual fast rotating massive stars computed in Paper I;
    the label indicates the corresponding initial stellar mass. The dotted
    line corresponds to ejecta of a 60~\Ms{} stars computed with nominal
    reactions rates (model 60rB in \citealt{DecressinMeynet2007}).  We use
    the minimum dilution factor $a_{min} = 0.3$.}
  \label{fig:Li1}
\end{figure}

\begin{table}[htbp]
  \caption{Pristine composition and mean composition of slow winds after
    convolution by an IMF for some elements, He is in mass fraction.}
  \label{tab:mix}
  \begin{tabular}{cccc}
    \hline 
    Element & Initial & \multicolumn{2}{c}{IMF slope (x)} \\
    & & 0.4 & 1.35 \\
    \hline
    \hline
    {[Na/Fe]}  & $-0.28$ &  0.47 &  0.44 \\
    {[O/Fe]}   &  0.32 &  0.06 &  0.10 \\
    {[N/Fe]}   & $-0.18$ &  0.99 &  0.94 \\
    {[C/Fe]}   & $-0.16$ & $-0.56$ & $-0.53$ \\
    $A($Li$)$  &   2.6 & $-\infty$ & $-\infty$\\
    He         & 0.245 &  0.40 &  0.37 \\
    \hline
  \end{tabular}
\end{table}

Let us now turn to the dilution that occurs between the slow ejecta of
massive stars and pristine gas.  If all the ejecta of all potential
polluters were fully mixed before the formation of the second generation
stars and no dilution with ISM occurs,
Eq.~(\ref{eq:dilution}) should be replaced by:
\begin{equation}
  \label{eq:fullmix}
  X_i^\text{2G} = \frac{ \int_{20}^{120}\tilde{m}_\text{SW} \tilde{X}_i
    \Phi(M) \diff M}
  {\int_{20}^{120}\tilde{m}_\text{SW} \Phi(M) \diff M},
\end{equation}
with, for a given individual star $\tilde{m}_\text{SW}$ and $\tilde{X}_i$
being respectively the total mass ejected in its slow wind and the mean
mass fraction of a given element $i$ in its ejecta.  In that case second
generation stars would all have the same composition that is given in
Table~\ref{tab:mix} for the two extreme IMF slopes found previously (i.e.,
0.4 and 1.35)\footnote{Note that the ejecta composition is weighted by
  pristine matter released by the polluter at the beginning of the main
  sequence, when rotational-mixing did not have the time to convey
  H-burning products from the convective core to the surface. For this
  reason the 60 and 120~\Ms{} stars eject half of their slow winds with
  only little abundance variations. This explains the very similar values
  given in Table~\ref{tab:mix} for the two different IMF slopes.}. If we
take this fully mixed material and then dilute it in a stochastic way with
pristine gas, one can form second generation stars with composition ranging
between that of pristine matter (in the case of very strong dilution) and
that of the mixed ejecta (in the case without dilution). Only a small range
of abundance variations\footnote{We use the two standard spectroscopic
  notations: $[\text{X/Fe}] = \log (\text{X/Fe}) - \log
  (\text{X}_\odot/\text{Fe}_\odot)$, and $A(\text{X}) = \log(\text{X/H})
  +12$ with X, Fe, and H in number density.} can be achieved in this way.
In this case, the theoretical [O/Fe] ratios vary by only
0.26~dex, whereas the observed [O/Fe] abundances span more than 1~dex in
NGC~6752 \citep{CarrettaBragaglia2007}.

As discussed in Sect.~2, there is additional compelling evidence that
the slow wind of an individual star pollutes only on a small scale around
its stellar progenitor, and that it is diluted locally with pristine ISM.
In order to constrain quantitatively this local dilution process we use the
Li variation detected in NGC~6752 by \citet{PasquiniBonifacio2005} (see
Fig.~\ref{fig:Li1}).
It is important to recall that the massive star ejecta are totally
``Li-free'', as this fragile element is destroyed in these objects.
We suppose that the matter ejected by an individual massive stars early on
the main sequence encounters more pristine gas and is more diluted than the
matter ejected later. As the winds are more and more enriched in Na in the
course of the evolution of a massive polluter, we expect an anticorrelation
between Li and Na: first, Li-free matter ejected with a low (i.e., close to
initial) Na content is more diluted in pristine matter which is Li-rich;
low-mass stars with a relatively high Li content and a relatively low Na
abundance are then created.
Later, Li-free matter ejected with high Na abundance is diluted with less
pristine material; as a consequence, the newly formed stars have less Li
while their Na is high.
The Li-Na anticorrelation observed by \citet{PasquiniBonifacio2005}
thus provides a calibration of the dilution factor $a$ in
Eq.~(\ref{eq:dilution}). In this paper we use the following expression for
this parameter:
\begin{equation}
  \label{eq:a}
  a_t = 1 - \left(1-a_\text{min}\right)
  \frac{t}{t_*}
\end{equation}
with $t$ and $t_*$ respectively the current time and the total lifetime
during which low-velocity winds are ejected. $a_\text{min}$ indicates the
lower value for the
dilution and $a$ decreases from 1 to $a_\text{min}$ when the polluter star
evolves.

Since Li is destroyed in massive stars, we have that
\begin{equation}
  \label{eq:dilution2}
  X_{\text{Li}}^{2G} (a_t) = a_t X_{\text{Li}}^\text{init},
\end{equation}
where $X_{\text{Li}}^{2G}$ is the mass fraction of Li in second generation
stars at birth, $X_{\text{Li}}^\text{init}$ is the value derived from WMAP
data and the standard Big Bang nucleosynthesis (A(Li)=2.65 according to
\citealp{Steigman2006}).  To take into account the Li depletion (of about
0.3~dex) occurring in low-mass stars \citep[see][]{CharbonnelPrimas2005},
the values of $X_{\text{Li}}^{2G}$ are lowered by 0.3~dex.

In Fig.~\ref{fig:Li1} we superimpose our theoretical tracks for the Li-Na
anticorrelation obtained with a value of $a_\text{min}$ equal to 0.3 on the
observed anticorrelation determined by \citet{PasquiniBonifacio2005} in
NGC~6752. All the tracks of massive stars show a clear Li-Na
anticorrelation with a small dispersion between those tracks (see the
continuous lines in Fig.~\ref{fig:Li1}).

The anticorrelation follows the same trend in our models and in turnoff
stars observed by \citet{PasquiniBonifacio2005} in NGC 6752. A dilution
factor, $a_\text{min}$, around $0.3$ is needed to reproduce the stars with
the lowest Li abundance. 
An overall agreement (within the observational errors) is obtained
  with this value as shown in Fig.~\ref{fig:Li1}. However it should be
  noted that uncertainties on the used nuclear reaction rates can modify
  this anticorrelation.
%
To
illustrate this point we show in Fig.~\ref{fig:Li1} the wind composition of
a 60~\Ms{} rotating star computed with a different set of reaction rates
(dotted line, model 60rB in paper I). In this case, the Na enrichment
largely exceeds the observed values for the same dilution factor. Thus the
uncertainties about the nuclear reaction rates prevent firmer
conclusions from being reached.

With the above prescription for $a_t$ we can use Eq.~(\ref{eq:globloc}) to
compute the global dilution factor $d$. We find that $d \simeq 1.15$ for
both scenarios. The previous findings on the mass budget obtained with
$d=1$ are thus still valid.

\subsection{Theoretical ONa distribution}

\begin{figure}[tbp]
  \includegraphics[width=0.5\textwidth]{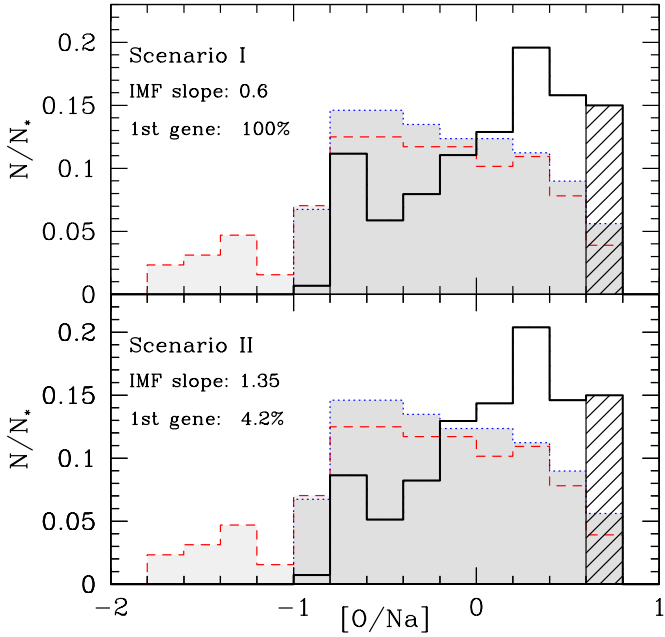}
  \caption{Distribution function of [O/Na] obtained in Scenario I
      (top) and II (bottom) shown as thick lines. Variable dilution factor
      with $a_\text{min} = 0.3$ is taken into account (see
      Eq.~\ref{eq:a}). The label ``1st gene'' indicates the percentage of
      first generation long-lived stars remaining in the cluster throughout
      its evolution. Hatched areas indicate the first generation or
      ``normal'' stars.  We also superimpose the observed distribution
      histogram in NGC~6752 \citep{CarrettaBragaglia2007} indicated with
      the dotted and dashed thin lines which have the same meaning as in
      Fig.~\ref{fig:ona6752}.}
  \label{fig:obsONa}
\end{figure}

We compute the theoretical [O/Na] distribution function with the parameters
we have set up previously for Scenarios~I and II (see Eqs.~\ref{eq:s1b} and
\ref{eq:s2b}). A variable dilution factor $a_t$ is used with the parameter
$a_\text{min} = 0.3$ (see Eq.~\ref{eq:a}). Our results are gathered in
Fig.~\ref{fig:obsONa} which displays the [O/Na] distribution function.
First generation stars representing 15\% of the present day distribution
are depicted in the hatched area in Fig.~\ref{fig:obsONa}.

Comparing the two upper panels of Fig.~\ref{fig:obsONa}, we see that the
distributions are strikingly similar. How are two such different scenario
possible? We first note that, by construction, the number ratio of
anomalous to normal stars is the same in the two scenarios (the parameters
have been chosen in order to obtain the same ratio of 85/15). One can make
this ratio appear in the denominator of the Eqs.~(\ref{eq:s1b}) and
(\ref{eq:s2b}) putting in evidence $f_\text{SW}(1+d)$. Doing this we see
that whatever the scenario considered, the distribution is proportional to
the ratio of two quantities involving only masses released by slow winds.
A decrease of the slope of the IMF thus produces a similar increase of both
the numerator and the denominator.

The peak in the distribution around $\text{[O/Na]} \sim -0.1$ is due to the
stars in the mass range 20-40~\Ms{} which have a longer evolution before
reaching the break-up, so that they are more mixed when they release their
slow winds.

The dilution with pristine gas adds to the slow winds material
characterised with high values of [O/Na], therefore dilution
increases the values of [O/Na]. With a $a_\text{min}$ parameter of 0.3 no
second generation stars can be produced with a [O/Na] ratio lower than -1
whereas large amounts of winds generated by the 60 and 120~\Ms{} have a
[O/Na] ratio lower than $-1$.  NGC~6752 can possibly harbour a population
of super-O-poor only detected with a strong Na abundance (see the dashed
line in Fig.~\ref{fig:obsONa}). For those stars the O abundance is not
directly measured but indirectly determined through the measure of the Na
abundance and using the Na-O anticorrelation. If this group is real we need
to change the minimal amount for the mixing to $a_\text{min} \simeq 0.03$.
But in that case, very low abundances of Li are expected. At the moment no
Li measurements have been performed at the surface of non-evolved stars with
such a high Na content. If feasible, such a measurement would represent an
interesting extension of the anticorrelation observed by
\citet{PasquiniBonifacio2005}.  Another possibility to explain these very
O-poor stars has been suggested by \citet{DAntonaVentura2007}. These
authors explain these stars as resulting from a mixing occurring in the
observed star itself along the red giant branch.  This mixing would occur
only in He-rich stars, because in these stars mixing would not be inhibited
by $\mu$-gradients.

If we compare our results with the distribution obtained by
\citet{CarrettaBragaglia2007} (lower panel in Fig.~\ref{fig:obsONa}) we can 
recover well stars with [O/Na] in the range between -1 and 0.7 although the
computed distribution does not recover the exact shape of that observed.
We however emphasise that the agreement is quite reasonable in view of
the very small number of free parameters. Indeed, given a set of fast
rotating massive star models and imposing a given value for the ratio of
anomalous to normal stars comparable to that observed in NGC 6752, the
computed distribution depends on only one parameter which is
$a_\text{min}$. Nature is probably more complicated. For example, we
considered only one rotational velocity for our massive stars, while real
stars probably present a distribution of velocities. Some uncertainties
pertain to the nuclear reaction rates (see the effect in Fig.~\ref{fig:Li1}),
which may affect the distribution of the composition of the slow winds and
therefore of the second generation stars. Moreover, as briefly indicated
above, the computed distribution depends on the dilution factor, a
parameter which is difficult to constrain precisely and which we modelled
here in a very simple way. In view of all these caveats, we think that the
results obtained are quite encouraging. In the next sections, we provide
some predictions which might be used in the future to confirm the
scenarios presented in this paper.

\section{Helium content, $^{12}$C/$^{13}$C and $^{16}$O/$^{17}$O isotopic
  ratios}

\subsection{He ejecta from massive polluters}

\begin{figure}[tbp]
  \includegraphics[width=0.5\textwidth]{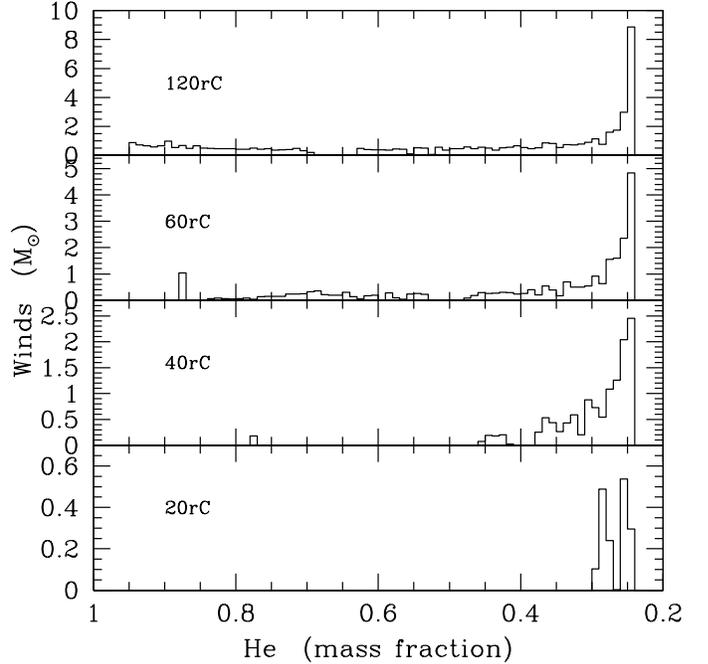}
  \caption{Mass ejected by slow winds with various helium contents for
    different initial mass models.}
  \label{fig:Heeject}
\end{figure}

Figure~\ref{fig:Heeject} shows the composition in helium (in mass fraction)
of the slow winds ejected by the massive star models of
\citetalias{DecressinMeynet2007}. The higher the initial mass of the
polluter, the higher the maximum He value in the wind. For models of 20
and 40~\Ms{} the mass fraction of He in the wind can reach a value
  of 0.30 and 0.45 respectively, while it can be as high as 0.80 in more
massive models.  Such a high value in the latter case is due to the
combined action of a more vigorous internal mixing during the evolution of
the polluter and especially during the main sequence, and to a stronger
mass-loss which reveals H-processed layers. For the 120~\Ms{} the He mass
fraction in the wind is already 0.80 at H-exhaustion and reaches 0.90 at
the end of the LBV phase.

\subsection{He content of second generation long-lived stars}
Let us now use these predictions for the He content of the slow wind of
massive stars, in order to derive the range in the He content of second
generation long-lived stars.

\begin{figure}[tph]
  \includegraphics[width=0.5\textwidth]{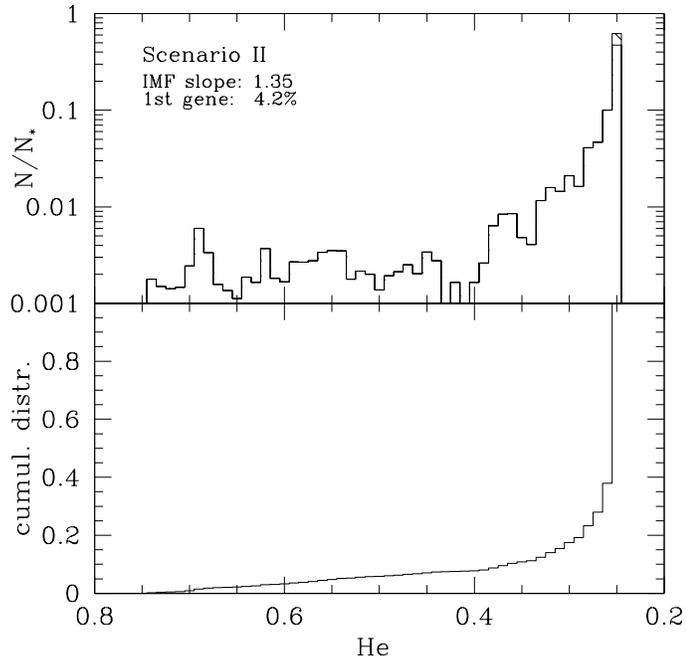}
  \caption{\emph{Top panel:} distribution function of He (mass fraction) of
    low-mass stars for second (white area) and first (hatched area)
    generation stars. \emph{Bottom panel:} cumulative distribution function
    of He in low-mass stars.}
  \label{fig:He0}
\end{figure}

The number fraction of second generation long-lived stars as a function of
their helium content is shown in Fig.~\ref{fig:He0}.  Only the case for
Scenario~II is shown (the results would be very similar in the case of
Scenario~I). The theoretical distribution is that expected at the birth of
the second generation stars\footnote{Internal mixing processes may indeed
  slightly change the surface composition of stars during their lifetime
  (for instance during the red giant phase at the first dredge-up the
  \el{C}{12}/\el{C}{13} drops to about 3.5, with C and Li decrease
  accompanied with an N increase). Thus the present predictions can only be
  compared with the observed surface composition of present day low mass
  stars which are still on the Main Sequence or at an evolutionary stage
  before the first dredge-up episode.}.  An extended tail is present toward
very high values (up to 0.75) with around 12\% of the low-mass stars
displaying a He content higher than 0.4.

It should be noted that some observational features in globular
  clusters such as the multiple sequences in $\omega$~Cen
  \citep{BedinPiotto2004} and NGC~2808 \citep{PiottoBedin2007} are
  explained by an He enrichment of the low-mass stars up to He $\simeq 0.4$
  \citep{Norris2004,Piotto2005,D'AntonaBellazzini2005}.  Differences in the
  HB morphology seem also related to high He value (around 0.4)
  \citep{D'AntonaCaloi2004,D'AntonaBellazzini2005,CaloiD'Antona2006,BussoCassisi2007}.
  Our model of fast rotating massive stars can easily produce winds with
  such a value and seem to be a good candidate to be at the origin of these
  features.  However we defer the discussion of this point to a future
  paper.

\subsection{Carbon isotopic ratio}

\begin{figure}[h]
  \includegraphics[width=0.5\textwidth]{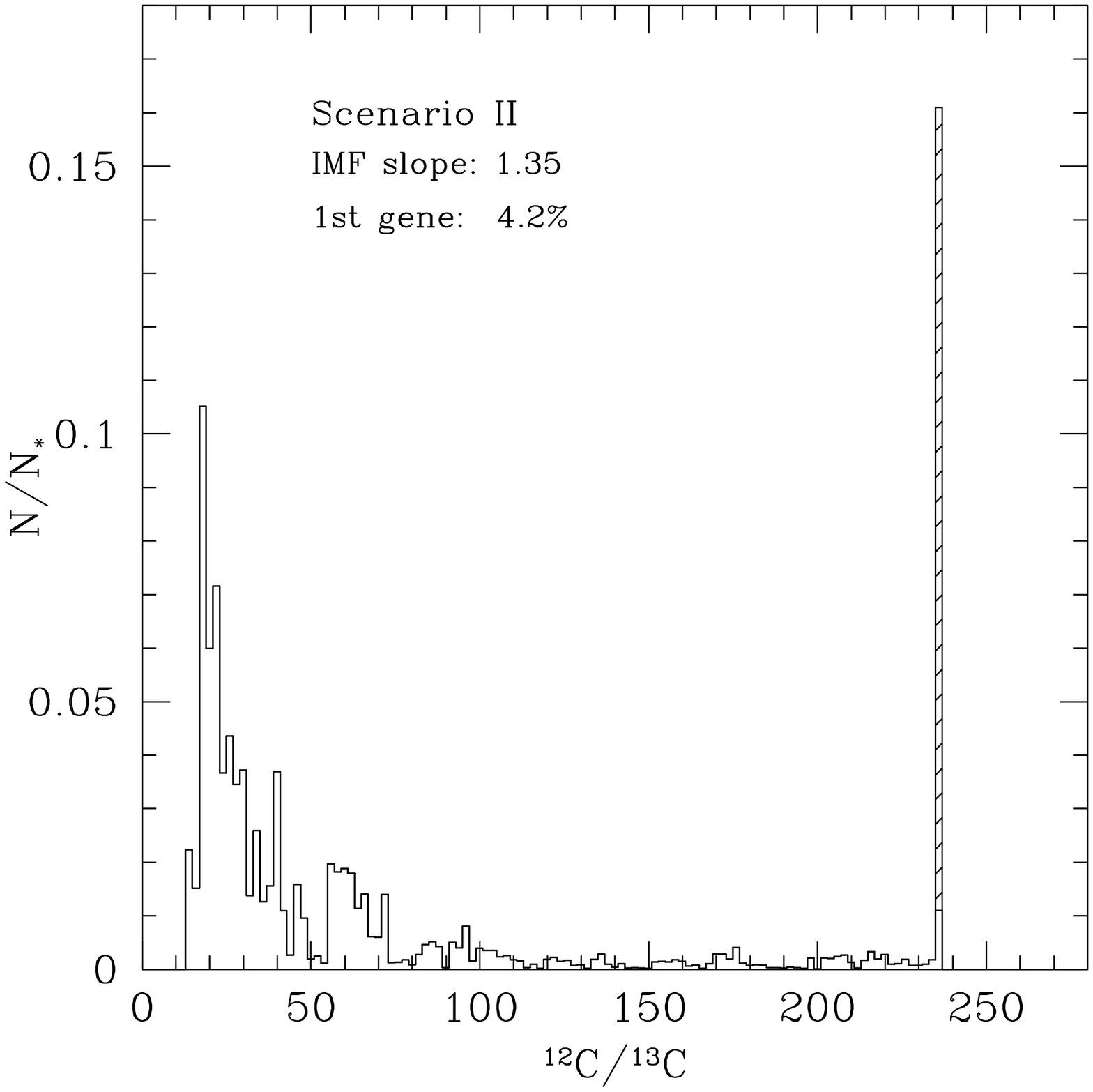}
  \caption{Number fraction of stars of the second generation composed of
    material with various \el{C}{12}/\el{C}{13} ratios. The hatched area
    corresponds to first generation stars.}
  \label{fig:YCiso}
\end{figure}
 
We do the same exercise for $^{12}$C and $^{13}$C. Figure~\ref{fig:YCiso}
shows the number fraction of stars as a function of their
\el{C}{12}/\el{C}{13} isotopic ratio at birth. Only the case with Salpeter
IMF (Scenario~II) is shown as the result is almost independent of the
adopted IMF. We obtain a bimodal distribution with a broader peak at low
\el{C}{12}/\el{C}{13} value. The peak around 240 is only made of first
generation stars\footnote{The value of 240 is taken from galactic chemical
  evolution from \citet{PrantzosAubert1996}. A different value around 90
  would not change the picture: we would still obtain a bimodal
  distribution with a high value narrow peak at 90 accompanied with a broad
  one at lower values.}. We note that most of the matter ejected in the
slow wind is near CN equilibrium, \ie{} it has \el{C}{12}/\el{C}{13} $\sim$
3.8. Dilution with pristine matter shifts the corresponding peak to
slightly higher values. Thus many second generation stars are expected to
be born with \el{C}{12}/\el{C}{13} around 15--20.

Some observations of the \el{C}{12}/\el{C}{13} are available in the
literature. For subgiants in NGC~6752, \citet{CarrettaGratton2005} found
values between 3 and 11 remarkably near the values we obtain for the peak
centered on low \el{C}{12}/\el{C}{13} values. Thus the ratio in second
generation stars is found, as expected, to be low but not at the
equilibrium. The observed values are however slightly smaller than the
theoretical one. This might indicate a lower dilution factor. Typically, to
retrieve the peak around 10, we need to use a dilution factor
$a_\text{min}$ of 0.1.

\subsection{Oxygen isotopic ratios}

\begin{figure}[tb]
  \includegraphics[width=0.5\textwidth]{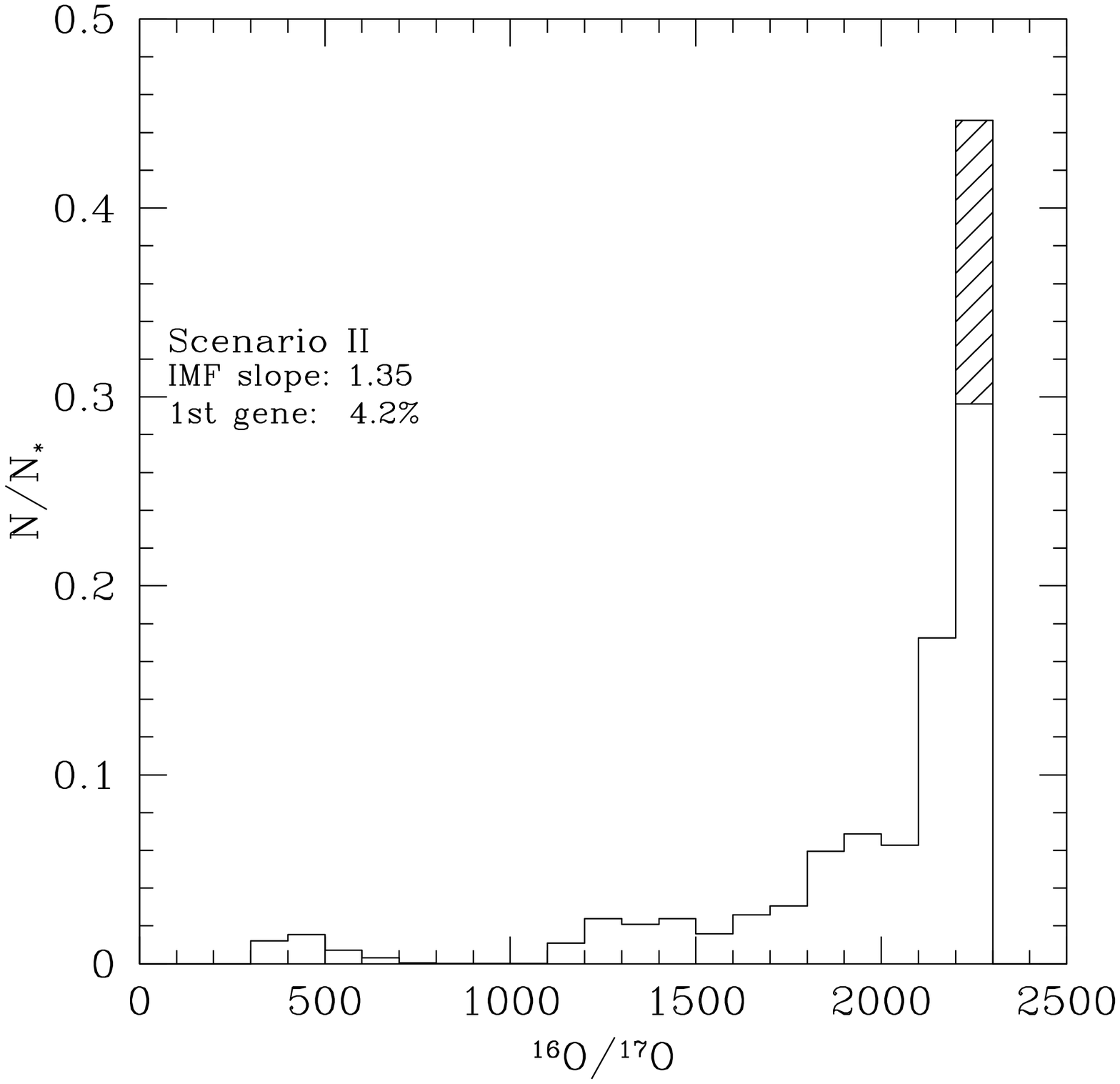}
  \caption{Number fraction of stars of the second generation composed of
    material with various \el{O}{16}/\el{O}{17} ratios. The hatched area
    corresponds to first generation stars.}
  \label{fig:YOiso}
\end{figure}

Let us first recall that, during the core H-burning phase, $^{17}$O is
produced in stars with initial mass $\le 40$~\Ms{} and is slightly
destroyed in more massive stars, while \el{O}{16} is destroyed in all
massive stars. Since in all cases, the burning rate of \el{O}{16} is faster
than that \el{O}{17}, the isotopic ratio\footnote{The nuclear rates of the
  reactions \el{O}{17}$(p,\gamma)$ and \el{O}{17}$(p,\alpha)$ present large
uncertainties for temperatures higher than $100 \times 10^6$~K which
undermine the predictions for explosive nucleosynthesis. However at
temperatures reached at the center of main-sequence massive stars (around
$50 \times 10^6$~K) those uncertainties remain small, at similar level than
that of \el{O}{16}$(p,\gamma)$.}
\el{O}{16}/\el{O}{17} decreases
from 2500 (original value) to values between 300 (20~\Ms{}) and 650
(120~\Ms{}).

Figure~\ref{fig:YOiso} displays the predictions for the initial composition
of anomalous stars obtained for Scenario~II with a variable dilution factor
($a_\text{min} = 0.3$). The gap between 400 and 1000 is merely artificial
as the lowest values are entirely due to the 20~\Ms{} while the 40~\Ms{}
stars start at a ratio of 1000. So the stars between 20 and 40~\Ms{} will
likely fill this gap. We see that only a few percent of stars present
\el{O}{16}/\el{O}{17} ratios of the order of a few hundred at birth.

\subsection{Expected (anti)correlations}

\begin{figure*}[tb]
  \includegraphics[width=0.49\textwidth]{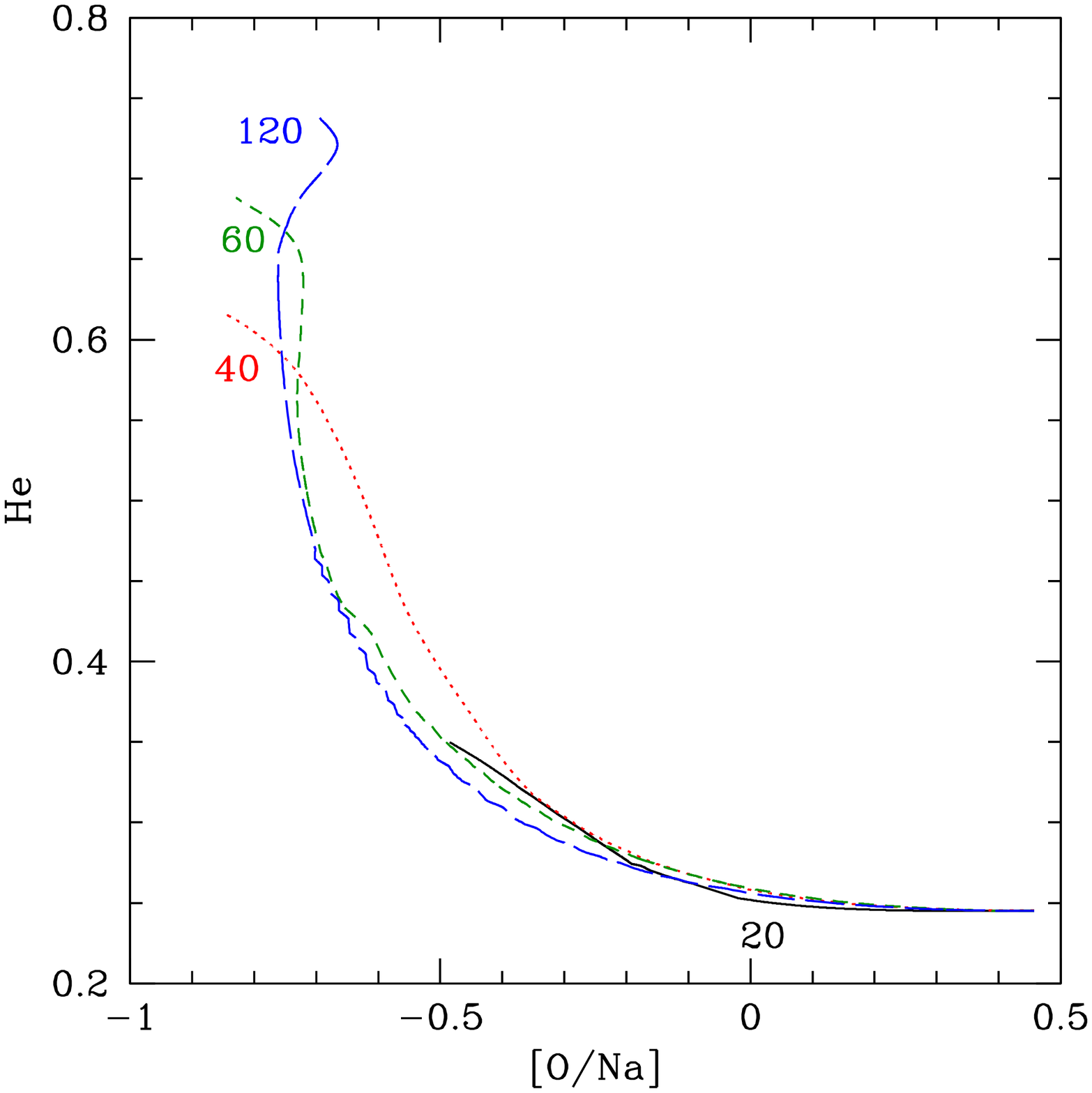}
  \includegraphics[width=0.49\textwidth]{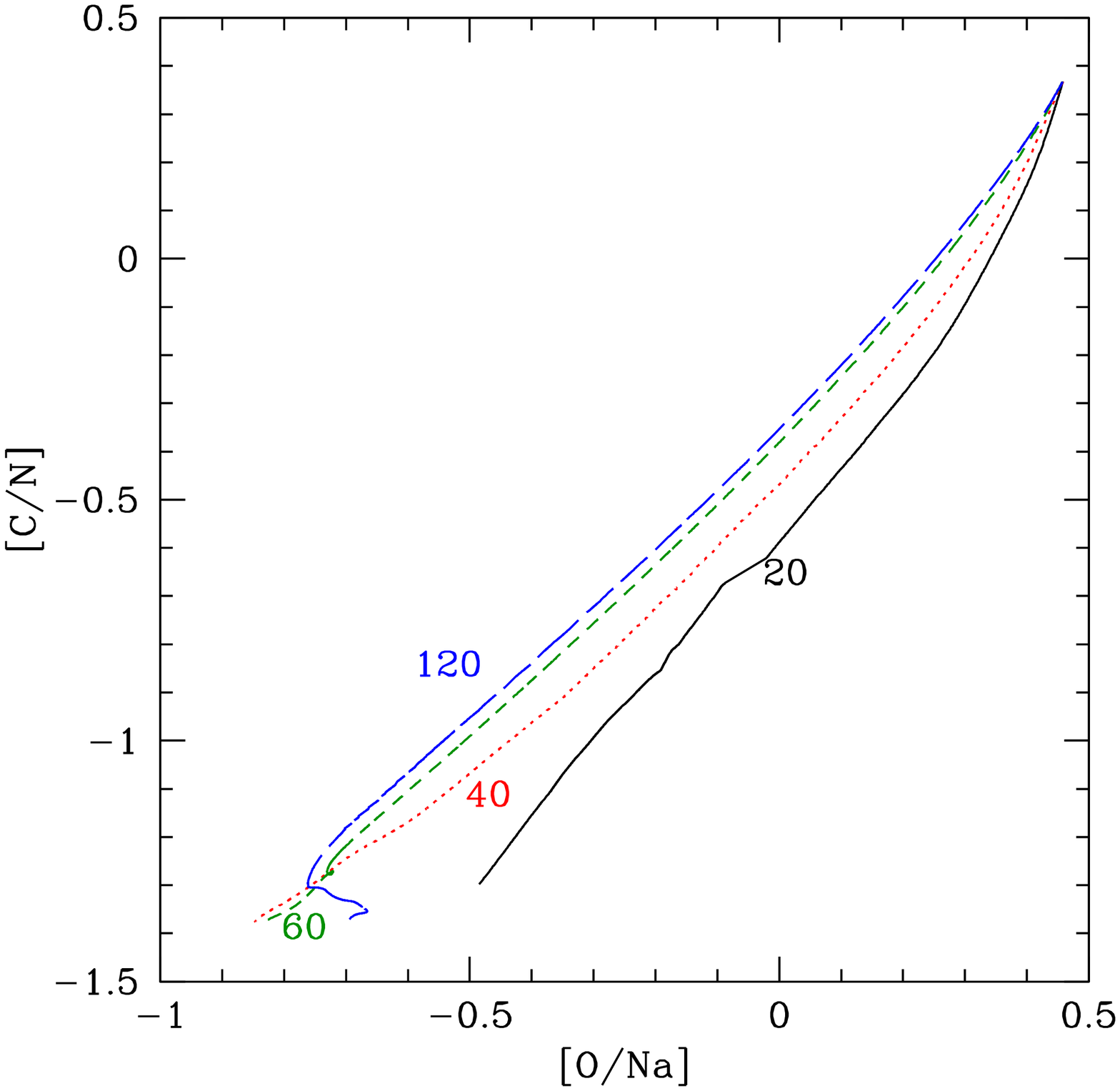}\\
  \includegraphics[width=0.49\textwidth]{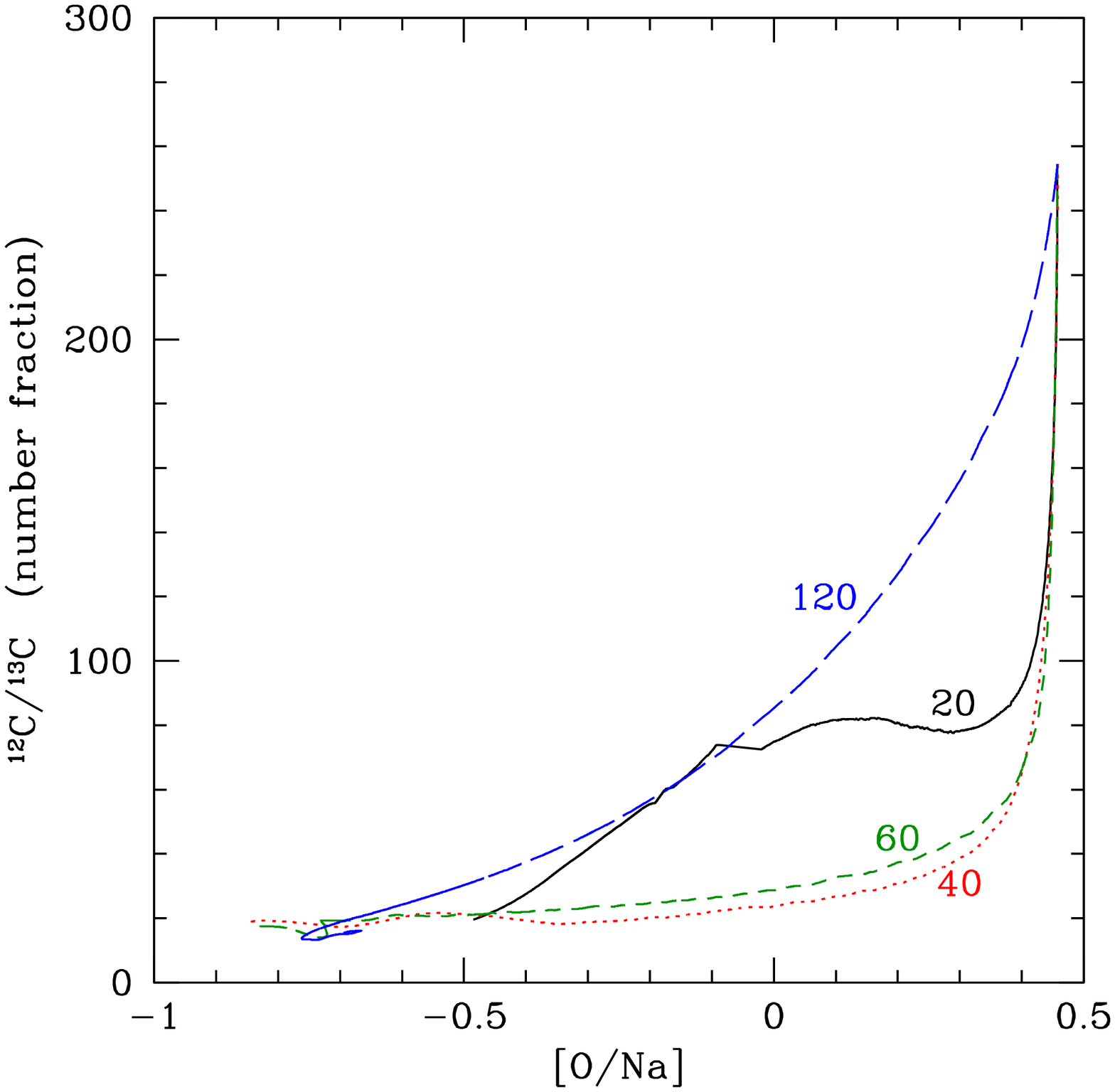}
  \includegraphics[width=0.49\textwidth]{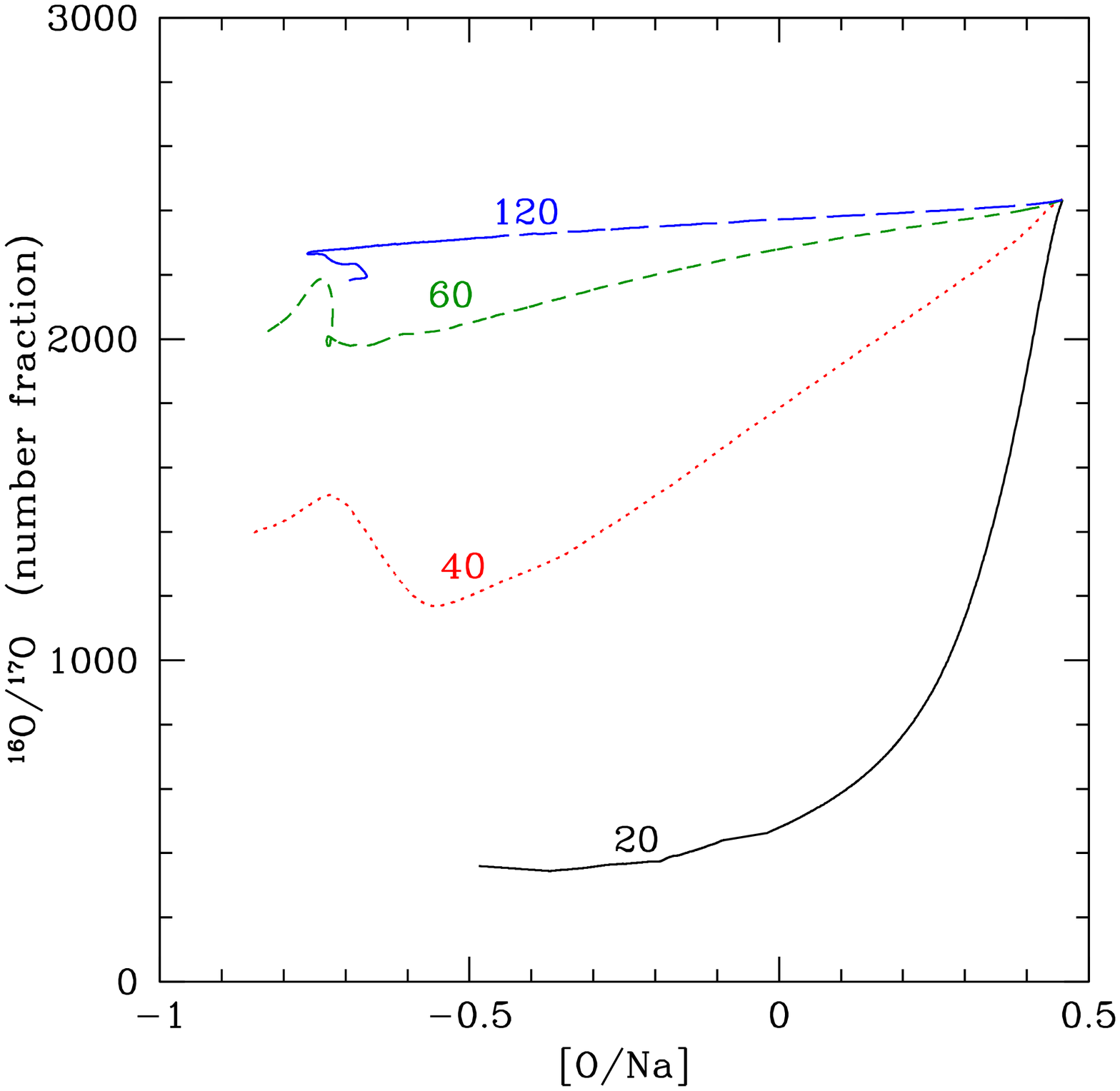}
  \caption{{\it Upper left:} Expected anti-correlation between helium and
    [O/Na] for second generation stars (at birth!) born from material
    ejected through slow winds of different initial mass models. {\it Upper
      right:} Same as the upper left panel for the correlation between
    [C/N] and [O/Na]. {\it Lower left:} Same as the upper left panel for
    the correlation between $^{12}$C/$^{13}$C and [O/Na]. {\it Lower
      right:} Same as the upper left panel for the correlation between
    $^{16}$O/$^{17}$O and [O/Na].}
  \label{fig:anticor}
\end{figure*}

Figure~\ref{fig:anticor} what are the expected relations between the
value of [O/Na] in second generation stars at birth and various other
chemical characteristics: helium content, [C/N], $^{12}$C/$^{13}$C and
$^{16}$O/$^{17}$O ratios. These (anti)correlations correspond to those
which should be observed in globular clusters today provided only
non-evolved stars are used, \ie{} stars which have not yet undergone
any change of their surface abundance due to internal mixing
processes. Many interesting points can be seen:
\begin{itemize}
\item The relations He-[O/Na] predicted by the different initial mass
  models present very small dispersions. This indicates that stars observed
  with a given value of [O/Na] should show very similar helium enrichment.
\item We see that the dispersion of the [C/N] ratio at a given value of
  [O/Na] is quite small if only the most massive stars are considered ($M
  \ge 40$~\Ms{}). Due to the 20~\Ms{} contribution, the dispersion can
  reach values of 0.4 dex in [C/N] at [O/Na] equal to about -0.5. In
  general, the higher the initial mass, the higher the [C/N] ratio for a
  given [O/Na] ratio. This reflects the fact noted above that the
  slow wind of lower initial mass stars contains mainly advanced processed
  material (low [C/N] values) because they reach the critical velocity only
  at a late time during the main-sequence phase.
\item The dispersion of the $^{12}$C/$^{13}$C ratio decreases for lower
  values of [O/Na].  All the stars with [O/Na] ratios below about -0.6 are
  predicted to show values of the $^{12}$C/$^{13}$C ratios between 15 and
  20.
\item The isotopic ratio $^{16}$O/$^{17}$O presents large dispersions for a
  given [O/Na] ratio on nearly the whole range of [O/Na] values. In the case
  that only very massive stars ($M\gtrsim 60$~\Ms{}) would contribute to
  provide the material rich in H-burning products, this ratio would remain
  above 2000.
\end{itemize}
It would be of course interesting to study to what extent
these relations depend also on the rotational velocities considered, on the
initial metallicity and on the mass range. It is of course the hope that
such predictions will stimulate new observations which will provide further
constraints on the origin of (part of) the matter which was used to form
the anomalous stars.

\section{Recall of the main results and discussion}

In this paper we have explored some consequences of the ``winds of fast
rotating massive stars'' scenario described in
\citetalias{DecressinMeynet2007}.

The two scenarios recalled above involve three free parameters: the
slope of the IMF of the polluters, the dilution factors (either the global
$d$, or the local $a_t$), and the number fraction of stars of the
first generation which has been lost by the cluster during its lifetime.

To constrain the values of these free parameters we used the
following observed features: the observed number ratio of anomalous to
normal stars and the Li-Na anticorrelation observed for turnoff
stars. Given a set of stellar models, this last feature depends only on the
local dilution factor $a_t$. From the observed Li-Na anticorrelation in NGC
6752 by \citet{PasquiniBonifacio2005} and adopting the
Eq.~(\ref{eq:dilution2}) for $a_t$, we obtain a value of 0.3 for
$a_\text{min}$, the minimum value of the dilution factor (see Sect.~5.5).
As expressed by the form of $a_t$, the mixing occurs in the vicinity of the
mass losing star. The degree of mixing varies as a function of the mass
losing star evolution. Inserting the expression for $a_t$ in
Eq.~(\ref{eq:globloc}), we obtain a global dilution factor $d$ of the order
of one. It means that the total mass of second generation stars amounts to
about twice the total mass released under the form of slow winds by massive
stars.

Given the above values for the dilution factors, the observed number ratio
of anomalous to normal stars depends only on the two remaining free
parameters. In Scenario~I, the cluster loses no stars and no stellar
remnants. In this case, the IMF is constrained. In Scenario~II, the
slope of the IMF is set to 1.35 and $e_\text{LL}^\text{1G}$ is
constrained. To produce a high number of polluted stars (typically 5.7
times the number of normal stars as observed in NGC 6752) we need either
(a) in Scenario~I, a flat IMF with a slope for massive stars around 0.55 or
(b) in Scenario~II, that about 96\% of the unpolluted low-mass stars has
been lost by the cluster.

Using the two sets of free parameters obtained in Scenarios I and II, we
produced predictions for the following features: the variation of the
number fraction of (non-evolved) long-lived stars as a function of their
surface values of [O/Na], of helium, of $^{12}$C/$^{13}$C and of
$^{16}$O/$^{17}$O, the relations between the [O/Na] ratios and the helium,
$^{12}$C/$^{13}$C, $^{16}$O/$^{17}$O and [C/N] ratios.  Predictions for the
mass of the present day globular clusters expressed as fractions of the
mass of the gas which has been used to form stars has also been obtained, as
well as how the present day mass is distributed among first and second
long-lived generation stars, and stellar remnants.

The following results have been obtained:
\begin{itemize}
\item The way the number fraction of stars varies as a function of the
  surface composition is similar in the two scenarios provided the adopted
  free parameters give the same number ratio of anomalous to normal stars
  and the same dilution factors.
\item In both scenarios, we can produce a [O/Na] anticorrelation in the
  range $-1 \le \text{[O/Na]} \le 0.62$. Quite remarkably, this range
  corresponds to that observed in NGC 6752 if one considers only the set of
  stars for which both measurements of oxygen and sodium are available. The
  possible super-O-poor stars, whose presence is indirectly inferred in
  some globular clusters, need a lower dilution coefficient or some
  additional mixing occurring in the star itself.
\item In both scenarios with our adopted free parameters, around 12\% of
  the low-mass stars are expected to display a He content higher than 0.4
  (in mass fraction). Note that He-enrichment is obtained keeping the sum
  C+N+O equals to its initial value as requested by the observations.
\item A tight anti-correlation is predicted between the helium abundance and
  the value of the ratio [O/Na] at the surface of turnoff stars. How this
  anti-correlation is modified by some internal mixing occurring during the
  ascent of the red giant branch will be studied in a forthcoming paper.
\item More than 30\% of stars are expected to have surface
  \el{C}{12}/\el{C}{13} ratios between 10 and 30.
\item The ratios $^{12}$C/$^{13}$C and [O/Na] are correlated.  All the
  stars with [O/Na] ratios below about -0.6 are predicted to show values of
  the $^{12}$C/$^{13}$C ratios between 15 and 20.
\item Only a few percent of stars present \el{O}{16}/\el{O}{17}
  ratios of the order of a few hundred at birth.
\item A correlation with very a large dispersion is predicted between the
  ratios \el{O}{16}/\el{O}{17} and [O/Na]. In the case that only stars more
  massive than about 60~\Ms{} would contribute to the building-up of the
  chemical anomalies in globular clusters, then only values above 2000
  would be expected for this ratio.
\item A correlation between [C/N] and [O/Na] is expected, whose dispersion
  increases at lower [O/Na] values.
\item In the case of Scenario I (IMF slope of 0.55,
  $e_\text{LL}^\text{1G}=e_\text{LL}^\text{2G}=e_\text{rem}^\text{1G}=e_\text{rem}^\text{2G}=0$,
  $a_\text{min}=0.3$), the present day mass cluster represents about half
  of the mass which has been used to form stars. The rest consists of
  matter ejected by stars and lost by the clusters.  The mass locked into
  remnants represents 19\% of the mass of the present day cluster, the
  first and second generation stars respectively 12 and 69\%.
\item In the case of Scenario II (IMF slope of 1.35,
  $e_\text{LL}^\text{1G}=e_\text{rem}^\text{1G}=0.958$,
  $a_\text{min}=0.3$), the present day mass cluster represents about 9\% of
  the mass which has been used to form stars. The rest consists of matter
  ejected by stars and of stars and stellar remnants lost by the cluster.
  The mass locked into remnants represents 8\% of the mass of the present
  day cluster, the first and second generation stars respectively 16 and
  76\%.
\end{itemize}

Obviously our two scenarios are schematic views of the real
  formation of globular clusters. We review their respective advantages and
  drawbacks in Sect.~7.2.

\subsection{Comparisons with other globular clusters}
\label{sec:discussion}

\begin{figure}[tb]
  \includegraphics[width=0.5\textwidth]{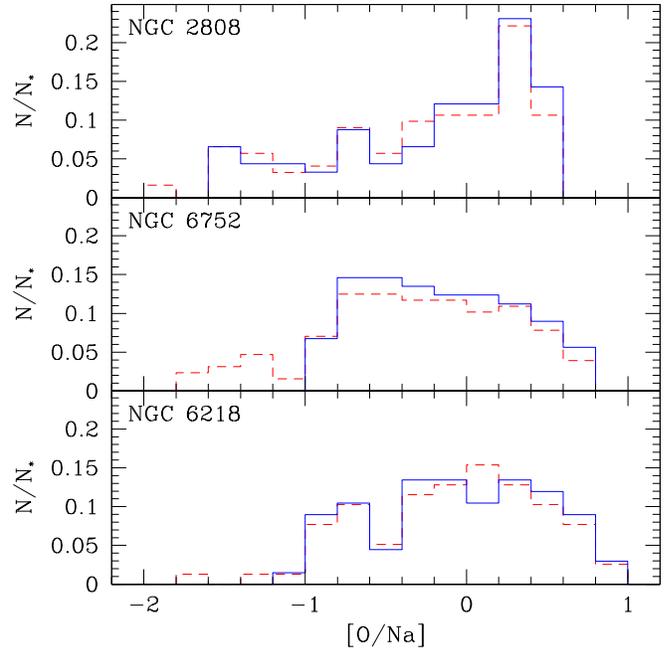}
  \caption{Observed [O/Na] distribution function in NGC~2808 (top),
    NGC~6752 (middle), and NGC~6218 (bottom). Full lines indicate
    distribution function obtained with detection or limits of O in stars
    while dashed lines indicate distribution obtained by a global
    anti-correlation relationship to obtain O abundance.}
  \label{fig:obsONa2}
\end{figure}

In the present study, we have provided some constraints on the slope of the
IMF, on the degree of evaporation of the cluster and on the dilution factor
based on observations of the globular cluster NGC 6752. We consider to
what extent using other globular clusters, the conclusion would have been
different.

First, let us recall the fact that the dilution factors have been
constrained on the basis of the Li-Na anticorrelation observed at the
surface of turnoff stars in NGC 6752. In the absence of similar data for
other clusters, we assume that the same dilution factor applies in
all globular clusters.

Thus for other globular clusters the other two observed feature which can
be used to constrain the slope of the IMF of the polluters and the degree
of evaporation of stars are the observed ratio of anomalous to normal stars
and the distribution of anomalous stars with given chemical abundances.

The observed ratio of anomalous to normal stars in \object{NGC 2808}
\citep{CarrettaBragaglia2006}, NGC~6752 \citep{CarrettaBragaglia2007}, and
\object{NGC 6218} \citep{CarrettaBragaglia2007b} are respectively 30\%
\citepalias{PrantzosCharbonnel2006}, 15\% and 15\%. As discussed in
Sect.~5.2, a higher proportion of first generation stars induces a steeper
IMF in both scenarios (see Fig.~\ref{fig:slope}). In the case of
Scenario~I, the IMF slope remains much flatter than a Salpeter IMF even
when the fraction of first generation stars amounts to 30\%. In the
framework of Scenario~II, a higher proportion of first generation stars
allows the cluster to lose less first generation stars, given an IMF slope.
However, passing from 15\% to 30\% for the fraction of first generation
stars will not change a lot the degree of evaporation. From
Fig.~\ref{fig:slope}, we can roughly estimate that it will cause the value
of $e_\text{LL}^\text{1G}$ decrease from 0.958 to about 0.85.

Figure~\ref{fig:obsONa2} displays the observed distribution function of the
[O/Na] ratio in NGC~2808, NGC~6752, and NGC~6218. Despite that the three
globular clusters have similar metallicity, they present distinct shapes
for their [O/Na] distribution function. In particular, NGC~2808 presents a
large population of super-O-poor stars. In NGC~6752 the presence of
super-O-poor stars is only inferred from the Na abundance, and NGC~6218 has
almost no such stars.  These differences are unlikely to be due to an
observational bias since the globular clusters have been studied with the
same protocol.  In fact they could reflect different dynamical histories
among the globular clusters. It should be noted that the present day
cluster mass of NGC~2808 is higher than that of NGC~6752, and NGC~6752 is
heavier than NGC~6218. This is in line with the finding of
\citet{Carretta2006} of indications that the GCs with higher
masses have more pronounced abundance variations.  With a similar IMF a
more massive cluster will have more massive stars available to create the
abundance anomalies. In addition we can expect that more massive clusters
have a deeper potential well and so a better ability to retain the slow
winds of massive stars.

From the above considerations, we can thus conclude that using other
globular clusters to constrain the free parameters of our two scenarios
would not significantly change our conclusions concerning the slope of the
IMF or the degree of evaporation of stars. However, probably other
parameters, such as the initial mass of the cluster, enter into play in the
global picture. We defer the discussion of this point to another paper
where further aspects of the dynamical history of the clusters
will be accounted for.

\subsection{The pros and cons of the two scenarios}

Finally we compare the two scenarios explored in this paper.

An advantage of Scenario~I is that it allows the cluster to retain its
first generation of stars and thus to suffer no or negligible
evaporation. Looking at the mass of young globular clusters, it seems a
reasonable hypothesis \citep{OstlinCumming2007}. Indeed these young globular
clusters present masses in the range of the observed masses of present day
old galactic globular clusters. Of course we are not sure that the young
globular clusters we see today are formed in the same way as those which
gave birth to the old galactic globular clusters. Most of the young
globular clusters are formed during galaxy merging, while at least part of the
old galactic globular clusters may have been formed by a rapid collapse a
proto-halo cloud. On the other hand a disadvantage of Scenario~I is that it
imposes a very flat IMF. At the moment there is no direct empirical
evidence that the IMF depends on any physical parameters such as for
instance star formation efficiency, the metallicity or the stellar
density \citep{MasseyJohnson1995}. Thus this is probably at the present
time a drawback of this scenario.

The advantage of Scenario~II is that it allows the use of a standard slope
for the IMF. The disadvantage however is that it imposes a very serious
loss of first generation stars. We propose a reasonable explanation for
why only first generation stars are lost based on the mass
segregation of massive stars, but we must admit that the requirement to
lose more than 95\% of this generation of stars might be too
demanding. \citet{BaumgardtKroupa2007} compute a grid of globular clusters
(N-body simulations) to study the effect of residual-gas expulsion on the
survival rate and final properties of star clusters. They found that some
survival clusters could lose a large part of their initial mass (up to
85\%). Models of the dynamics of globular clusters with two stellar
populations are still required to check whether the strong requirement we
need can be achieved or not \citep[\eg{}][]{DowningSills2007}.

Thus, we see that whatever scenario is realised, the peculiar abundance
patterns observed in globular clusters requires some quite extreme
assumptions: either a flat IMF or a very strong evaporation of stars or a
combination of the two aspects. Moreover, as explained in more detail in
paper I, we have to assume that the first generation of massive stars
contained a very high proportion of fast rotators.  Is that high proportion
of fast rotators due to the peculiar mode of star formation in dense
clusters? Are there any other observational hints supporting the existence
of these fast rotators? What would be the consequences of this population
of massive fast rotators on the frequency of gamma ray bursts in very young
globular clusters?  We leave open these questions for the moment and we
shall probably come back to them in future work.  A point however which
appears worthwhile to underline here is that massive fast rotators appear to
be interesting objects not only to explain the abundance patterns in
globular clusters, but also the high N/O ratios observed in very metal poor
halo stars \citep{ChiappiniMatteucci2005,ChiappiniHirschi2006} and some
C-rich Ultra Metal Poor Stars \citep{MeynetEkstrom1006,Hirschi2007}.

\begin{acknowledgements}
  The authors wish to thank P. Kroupa and H. Baumgaard for fruitful
  discussion concerning Scenario II, as well as N. Prantzos. We acknowledge
  the financial support of the Swiss National Science Foundation (FNS) and
  of Programme National de Physique Stellaire (PNPS) of CNRS/INSU, France.
\end{acknowledgements}

\bibliographystyle{aa}
\bibliography{BibADS}

\end{document}